\documentclass[conference]{IEEEtran}
\IEEEoverridecommandlockouts
\usepackage{cite}
\usepackage{amsmath,amssymb,amsfonts}
\usepackage{algorithm}      
\usepackage{algpseudocode}  

\usepackage{amsthm}
\newtheorem{example}{Example}
\newtheorem{Definition}{Definition}

\usepackage{tikz}
\usetikzlibrary{petri, positioning}

\usepackage{graphicx}
\usepackage{textcomp}
\def\BibTeX{{\rm B\kern-.05em{\sc i\kern-.025em b}\kern-.08em
    T\kern-.1667em\lower.7ex\hbox{E}\kern-.125emX}}

\usepackage{multirow}
\usepackage{caption}
\usepackage{url}
\usepackage{hyperref}
\usepackage{algorithm} 
\usepackage{listings}  
\usepackage{amsfonts}
\PassOptionsToPackage{most}{tcolorbox}
\usepackage[most]{tcolorbox}
\definecolor{blockGray}{gray}{0.95}
\usepackage{hyperref}
\newtcolorbox{myquote}{colback=blockGray, grow to right by=-4.8mm, grow to left by=-4.8mm, boxrule=0.7pt, boxsep=0.7pt, top=0mm, bottom=0mm}
\usepackage{tikz}
\usetikzlibrary{arrows.meta, positioning}
\usepackage{booktabs}
\usepackage{mathrsfs}
\usepackage[numbers]{natbib}

\tcbuselibrary{listingsutf8,skins}
\newtcblisting{mycode}{
  colback=white!97!black,
  colframe=black,
  listing only,
  left=0pt,
  enhanced,
  top=0pt,
  bottom=0pt,
  boxrule=0.2pt,
  boxsep=0pt, 
  listing options={
    basicstyle=\scriptsize\ttfamily,
    breaklines=true,
    breakatwhitespace=true,
    tabsize=1
  }
}

\begin{document}

\title{Enhancing Uncertainty Quantification for Runtime Safety Assurance Using Causal Risk Analysis and Operational Design Domain}

\author{\IEEEauthorblockN{Radouane Bouchekir, Michell Guzman Cancimance}
\IEEEauthorblockA{fortiss GmbH, An-Institut Technische Universität München, Guerickestraße 25 \\80805 München, Germany \\
\{bouchekir, guzman\}@fortiss.org}
}

\maketitle

\begin{abstract}
Ensuring the runtime safety of autonomous systems remains challenging due to deep learning components' inherent uncertainty and their sensitivity to environmental changes. In this paper, we propose an enhancement of traditional uncertainty quantification by explicitly incorporating environmental conditions using risk-based causal analysis. We leverage Hazard Analysis and Risk Assessment (HARA) and fault tree modeling to identify critical operational conditions affecting system functionality. These conditions, together with uncertainties from the data and model, are integrated into a unified Bayesian Network (BN). At runtime, this BN is instantiated using real-time environmental observations to infer a probabilistic distribution over the safety estimation. This distribution enables the computation of both expected performance and its associated variance, providing a dynamic and context-aware measure of uncertainty. We demonstrate our approach through a case study of the Object Detection (OD) component in an Automated Valet Parking (AVP).
\end{abstract}

\begin{IEEEkeywords}
Uncertainty quantification,  Causality analysis,  Bayesian Network, Safety assurance.
\end{IEEEkeywords}

\section{Introduction}
The integration of Deep Neural Networks (DNNs) in autonomous systems (AS), such as Automated Valet Parking (\texttt{AVP}) \cite{esen2023simulationbased,sorokin2024towards}, has significantly advanced perception and decision-making capabilities. However, these learning-enabled components introduce substantial uncertainty and pose challenges for traditional, static safety assurance methods. This uncertainty arises not only from the model and data but also from dynamic, partially observable environments where safety assumptions may no longer hold at runtime. As a result, static safety assurance can quickly become obsolete, undermining trust in the system’s safe operation  \cite{denney2015dynamic,koopman2024redefining}.

To ensure trustworthy behavior, safety assurance must evolve from a static, design-time process into a dynamic, runtime capability that accounts for real-world operational variability. In particular, environmental conditions, such as lighting, weather, and road surface states, can significantly impact the perception and control components of AS, especially those powered by DNNs. Therefore, quantifying how such conditions influence model behavior and propagating their effects into safety arguments is essential for robust runtime assurance \cite{shakeri2024operational}.

In this paper, we propose a methodology that enhances traditional uncertainty metrics by explicitly incorporating environmental conditions into the confidence estimation process. Our approach combines Hazard Analysis and Risk Assessment (HARA) with causal modeling techniques to identify environmental and contextual conditions that affect the functionality of safety-critical components. These relationships are encoded in a Bayesian Network (BN) that unifies uncertainty from the model, the data, and the operational context. At runtime, this BN is instantiated using real-time environmental observations, enabling the computation of a posterior distribution over the safety estimate. This distribution yields both the mean estimate (e.g., likelihood of correct behavior) and its variance (uncertainty), providing a dynamic, context-sensitive confidence metric.

A key contribution of our approach lies in its formalization strategy. To support machine-interpretable, consistent, and verifiable representations of the Operational Design Domain (ODD), safety arguments, and confidence models, we adopt an ontological framework grounded in Description Logic (DL). This formal representation facilitates automated reasoning, semantic querying, and traceability across assurance artifacts. Furthermore, by aligning our ontology with the ASAM OpenODD \cite{asam2021openodd}, we ensure semantic coherence and interoperability with emerging safety engineering practices. Specifically, the main contributions of this work are as follows: \begin{itemize} 
    \item[-] Systematic methodology for defining and formalizing ODD. 
    \item[-] Confidence model creation that explicitly incorporating environmental conditions into the confidence estimation process. 
    \item[-] Runtime monitoring mechanism that dynamically computes confidence scores using instantiated ODD attributes. 
\end{itemize}

In the following sections, we structure our work
as follows:  Section~\ref{sec:AVPUC} introduces the \texttt{AVP} case study. Section~\ref{sec:Methodology} details our methodology for quantifying uncertainty. Section~\ref{sec:ODD} presents the process for ODD determination and formalization, and Section \ref{subsec:safety-assurance} illustrates the safety argument pattern and their formalization. Section~\ref{sec:Confidence_Model} explains the creation and formalization of confidence models. Section~\ref{subsec:operation-time-confidence} demonstrates runtime confidence computation, and Section~\ref{sec:Implementation} illustrates the implementation details.  We conclude in Section~\ref{sec:conclusion}.

\section{The \texttt{AVP} Case Study} 
\label{sec:AVPUC}

This section presents a case study of the Automated Valet Parking (\texttt{AVP}) system, developed as part of the FOCETA project\footnote{\url{https://www.foceta-project.eu/}}.  The \texttt{AVP} functionality enables a vehicle to autonomously navigate to and park in an available parking space without human intervention \cite{BoschAVP}. In a typical scenario, the driver leaves the car in a designated handover area, after which the \texttt{AVP} system assumes control, computes a safe trajectory, and autonomously drives the car to a free parking spot.

\begin{figure}[!h] \centering \includegraphics[scale=0.33]{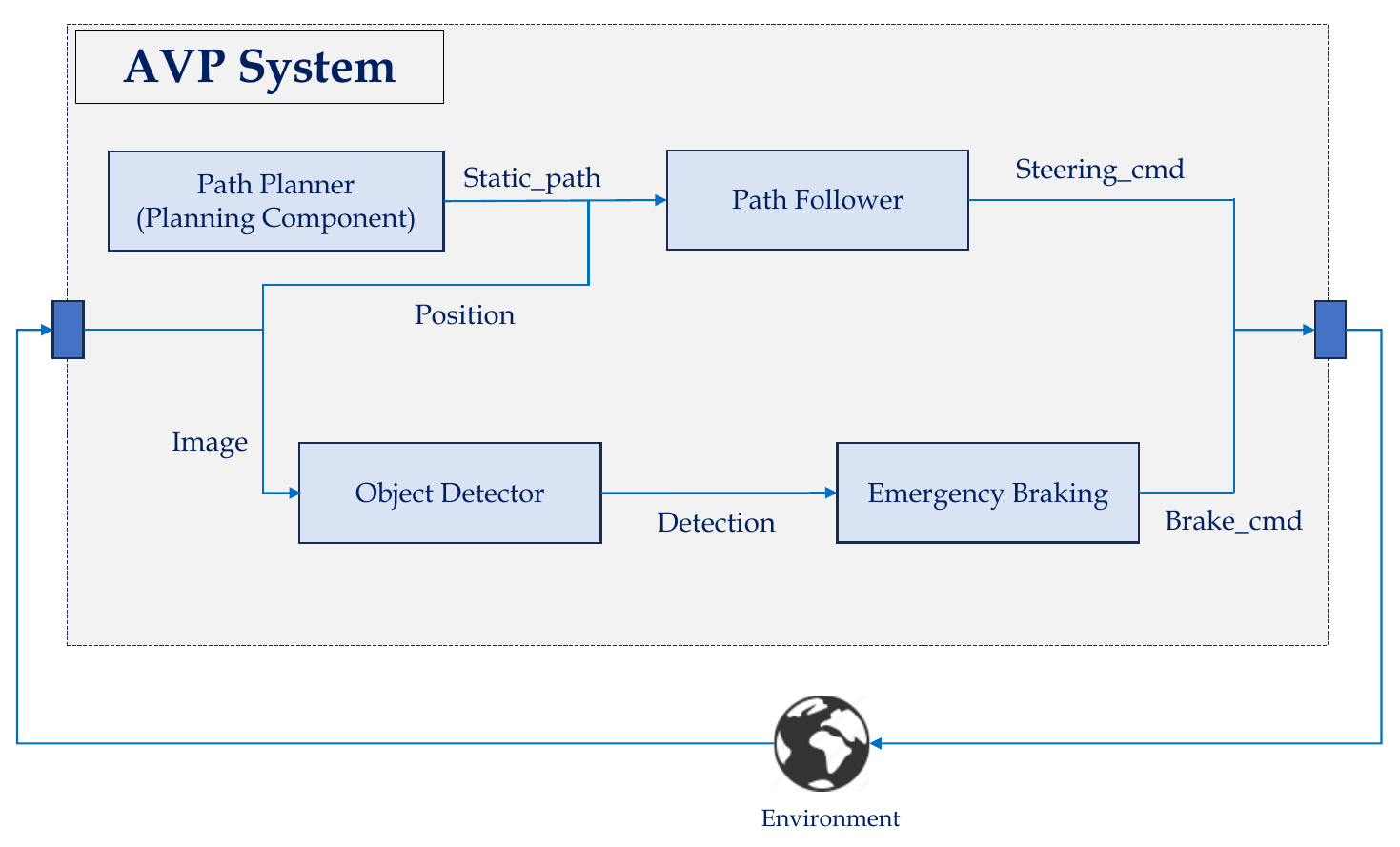} \caption{System architecture of the Automated Valet Parking (\texttt{AVP}) system.} \label{fig:avp-sut} \end{figure}

The \texttt{AVP} system architecture, depicted in Fig.~\ref{fig:avp-sut}, is composed of multiple interacting components. The \textit{Planning Component} (\texttt{PC}) is responsible for generating a collision-free path from the drop-off location to the target parking spot. The \textit{Path Follower} (\texttt{PF}) ensures the vehicle adheres to this predefined path during navigation. To enhance safety, an \textit{Emergency Braking} mechanism is activated whenever the \textit{Object Detector} (\texttt{OD}) identifies an obstacle within a critical distance. The PF and emergency brake collectively constitute the system's \textit{Control Component} (\texttt{CC}).\\

\textbf{\textit{Challenges in Safety Assurance.}}
To ensure the safe operation of the \texttt{AVP} system, it must adhere to the following key safety requirement~\cite{esen2023simulationbased,sorokin2024towards}:
\begin{quote} 
\textit{``Safe\_OD Req: The ego vehicle shall not collide with pedestrians, unless its velocity is zero."} 
\end{quote}
However, establishing and maintaining a convincing safety argument that demonstrates compliance with \texttt{Safe\_OD} is non-trivial. The dynamic and uncertain nature of the operational environment, combined with the inherent limitations of DNNs, poses substantial challenges in guaranteeing safe behavior across all operating scenarios. While a safety case for the \texttt{AVP} system was proposed in \cite{sorokin2024towards}, it lacks a quantitative confidence argumentation framework to rigorously assess the trustworthiness of the claims made.
\section{Methodology}
\label{sec:Methodology}
\begin{figure*}[!ht]
    \centering
    \includegraphics[width=0.9\linewidth]{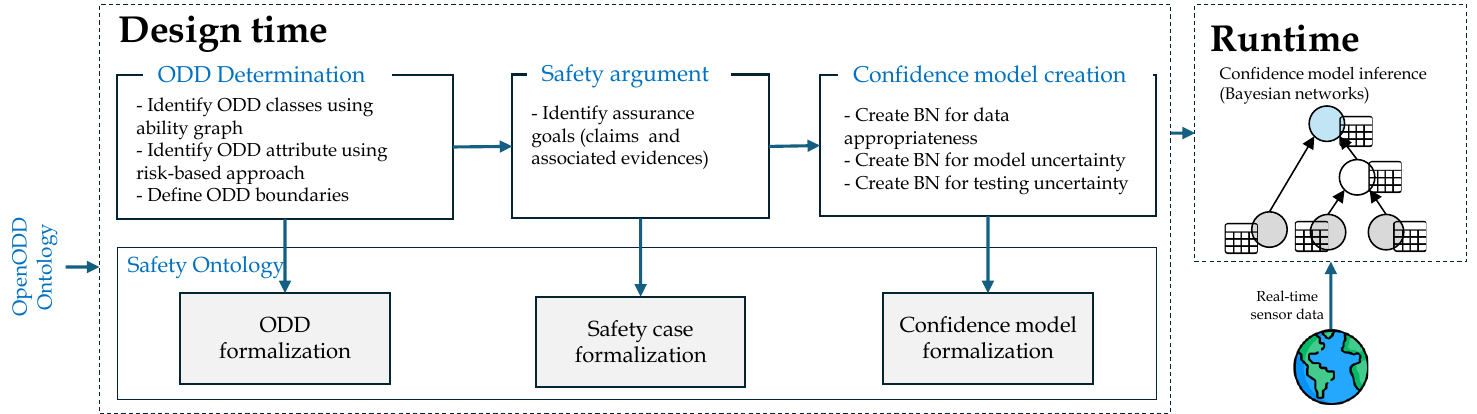}
    \caption{Overview of the proposed approach for DNN uncertainty quantification.}
    \label{fig:overiew_methodology}
\end{figure*}

The goal of our methodology is to construct a confidence-aware safety argument that accounts not only for the model and data but also for environmental variability. To achieve this, we quantify uncertainty in DNNs by integrating model, data, and environmental uncertainties within a unified BN-based framework. An overview of this methodology is presented in Fig.~\ref{fig:overiew_methodology}. The design phase begins with the determination of ODD classes and attributes. This is accomplished through a combination of ability graphs, which capture dependencies between system capabilities and environmental conditions, and risk-based causal analysis. The analysis yields a causal structure that links hazards to specific, measurable operational conditions (e.g., fog, low illumination, surface wetness). These conditions are refined into ODD attributes with clearly defined operational boundaries. The formalization of the resulting ODD is done using DL in alignment with the OpenODD ontology\footnote{\url{https://www.asam.net/standards/detail/openodd/}}.

To support safety argumentation for DNN-based components, we adopt structured safety case patterns \cite{wozniak2020safety,hawkins2021guidance,sorokin2024towards}. These patterns define assurance goals, represented as claims and supported by corresponding evidence. While such safety arguments are typically static, we extend this process by constructing a confidence model that allows quantification of the trustworthiness of the supporting evidence. This model is realized as a BN constructed from the ODD definitions and risk analysis results, and it captures uncertainties related to data appropriateness, model robustness, and testing adequacy. To ensure a shared understanding among stakeholders and enable formal reasoning, we extended the ODD formalization with the safety case and confidence model formalization, resulting in a uniform \textit{Safety Ontology}. 

At runtime, the system continuously ingests real-time sensor data (environment measurement),  and computes a confidence score that reflects the DNN current level of safety assurance. This dynamic evaluation supports context-aware decision-making and aligns with industry safety standards such as ISO 26262 \cite{iso26262} and ISO DIS 21448 (SOTIF) \cite{sotif}.

In the following sections, we will illustrate our proposed methodology through a case study involving the \texttt{AVP} system (cf. Section~\ref{sec:AVPUC}), with a specific focus on constructing the confidence model for the \texttt{OD} component.

\section{ODD Determination and Formalization}
\label{sec:ODD}
\subsection{ODD Determination}
In the context of the \texttt{AVP} system, DNNs are primarily employed within the perception pipeline, notably in the \texttt{OD} component. The performance of these DNNs is highly sensitive to variations in environmental conditions, making their predictions unreliable in scenarios that deviate from the training distribution. Consequently, it is critical to formally define the ODD to delineate the boundary conditions under which the \texttt{AVP} system is expected to function safely and reliably. According to SAE J3016 (2018), the ODD refers to: 
\begin{quote}
    \textit{``Operating conditions under which a given driving automation system or feature thereof is specifically designed to function, including, but not limited to, environmental, geographical, and time-of-day restrictions, and/or the requisite presence or absence of certain traffic or roadway characteristics"}.
\end{quote}

While SOTIF further refines this concept as: 
\begin{quote}
    \textit{``Specific conditions under which a given driving automation system is designed to function"}.
\end{quote}
In the case of the \texttt{AVP} system, these definitions are crucial for specifying the environmental and contextual assumptions that underpin the safe operation of the system, particularly the \texttt{OD} component.

\subsubsection{ODD Classes Determination}
To derive the initial ODD, i.e., ODD classes that specify the high-level categories of environmental conditions relevant to the safe operation of the \texttt{AVP} system, we utilize \textit{ability graphs} \cite{reschka2015ability}, a hierarchical modeling approach that captures the dependencies between system/component-level abilities and external operational conditions. Originally developed to model guidance behavior in automated vehicles, ability graphs are particularly effective during early design phases for identifying areas where uncertainty may influence system performance. The graph is structured hierarchically: \textit{high-level system abilities} (e.g., \textit{A2: Plan dynamic path}) are decomposed into \textit{intermediate} and \textit{low-level abilities} that depend on sensor data and environmental conditions. We refer to these environmental conditions as \textit{ODD classes}. The key elements of an ability graph include:

\begin{itemize}
    \item \textbf{\textit{Ability Nodes:}} These represent functional capabilities of the system/component (e.g., \textit{``detect obstacle,''} \textit{``estimate obstacle distance''}) and capture the input dependencies required to fulfill these capabilities (e.g., sufficient lighting, visibility, and sensor fidelity).

    \item \textbf{\textit{Quality Requirement Edges:}} These directed edges connect higher-level ability nodes to their subordinate abilities, indicating functional dependencies within the graph. Each edge encapsulates a set of quality-related constraints that must be satisfied by the subordinate node to ensure the reliable realization of the higher-level ability. These requirements, such as timing constraints, sensing precision, or fault tolerance, can be derived from the system's functional requirements, particularly those defined during the concept phase of the ISO26262 development lifecycle.

\end{itemize}

\begin{example}
    Fig. \ref{fig:AbilityGraphAVP} illustrates the ability graph corresponding to the \texttt{AVP} use case. Each node in the graph represents a distinct system ability, while edges capture the dependencies on specific environmental and operational conditions. For instance, the ability \texttt{A8: Detect obstacle} depends on multiple environmental factors (ODD classes), including \texttt{traffic structure and infrastructure}, \texttt{weather conditions}, \texttt{sun angle}, and \texttt{illumination levels}.  To facilitate a structured and semantically meaningful representation, we align the environmental dependencies of ability nodes with corresponding ODD classes defined in the ASAM OpenODD ontology. 
    
    \begin{figure}[!h]
        \centering
        \includegraphics[width=\linewidth]{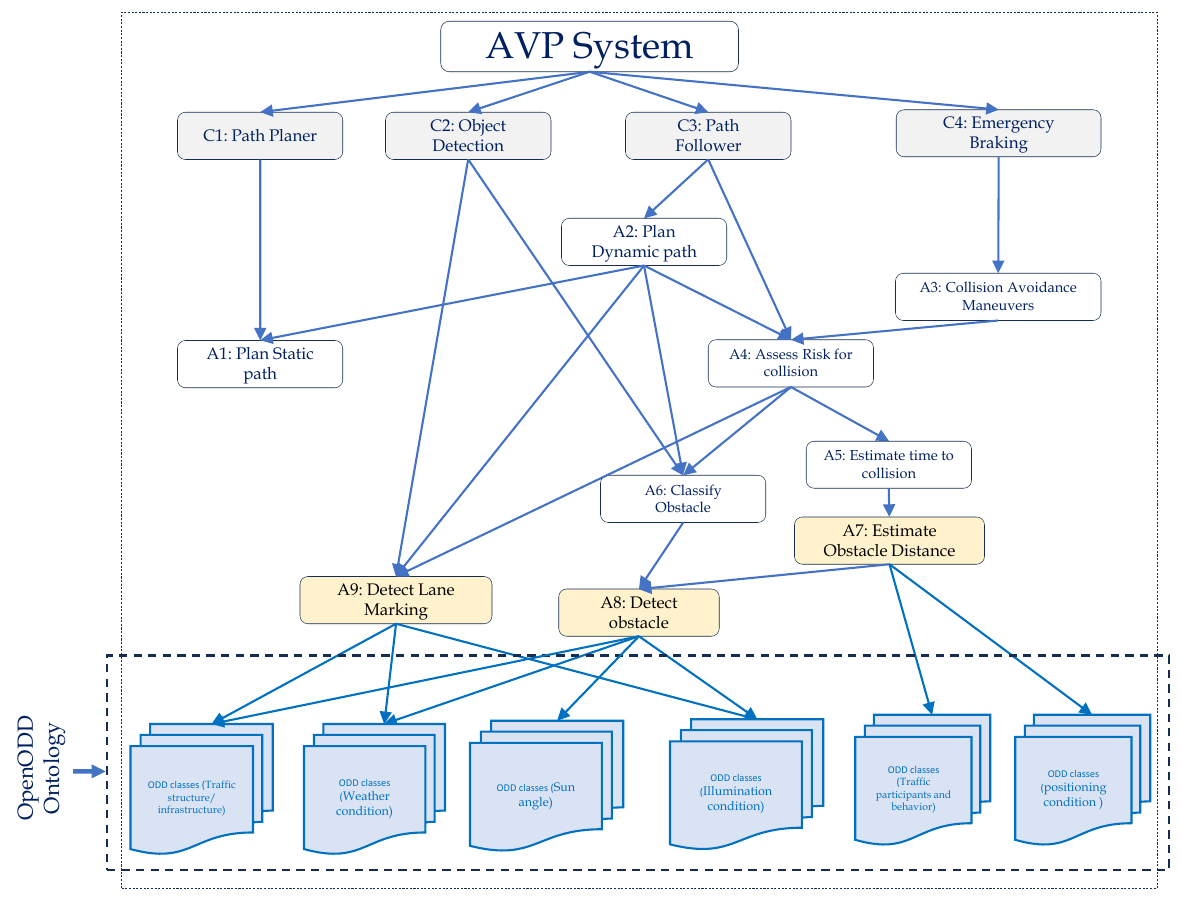}
        \caption{Ability graph for the \texttt{AVP} system, linking system abilities to relevant ODD classes.}
        \label{fig:AbilityGraphAVP}
    \end{figure}
\end{example}

\subsubsection{Risk-Based ODD Attribute Determination}
\label{subsubsec:risk-based-determination}
While the ability graph provides a structured way to identify ODD classes, it is essential to refine these classes by associating them with specific and measurable \textit{ODD attributes}. An \textit{ODD attribute} defines a concrete property (e.g., \texttt{rain intensity} or \texttt{fog density}) that adds granularity to the ODD class and enables precise specification of operational boundaries.

\begin{figure}[!h]
    \centering
    \includegraphics[width=\linewidth]{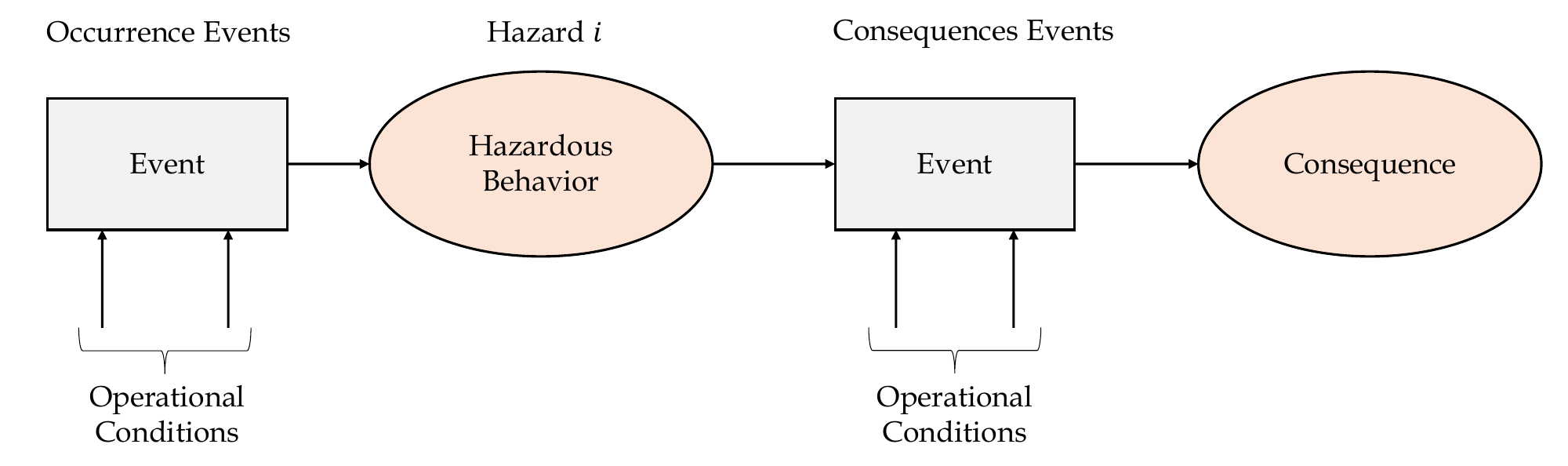}
    \caption{Causal chain illustrating occurrence and consequence events related to a hazard.}
    \label{fig:causal_chain}
\end{figure}

To derive these attributes systematically, we propose a \textit{risk-based} approach for ODD attribute determination. This approach builds upon established practices in the literature that emphasize the importance of risk assessment in defining the ODD \cite{louhichi2023new, cho2020operational, gyllenhammar2020towards, lee2020identifying, sun2021acclimatizing, macher2015sahara}. As illustrated in Fig. \ref{fig:causal_chain}, our methodology begins with the identification of potential hazards, defined as unreliable behaviors of the \texttt{AVP} system that may lead to a failure of its intended function. Once a hazard is established, we analyze its causal structure by identifying two types of related events:
\begin{itemize}
    \item \textbf{\textit{Occurrence Events:}} Events that lead to the manifestation of the hazard, often triggered by specific operational conditions (e.g., poor visibility or high speed).
    \item \textbf{\textit{Consequence Events:}} Events that follow the hazardous behavior and result in adverse outcomes (e.g., collision or injury).
\end{itemize}

Each of these events is linked to operational conditions. In fact, these events serve as the foundation for identifying relevant ODD attributes that contribute to or mitigate risk. This structured risk-based derivation process ensures that the defined ODD attributes are both operationally meaningful and safety-critical for the system/component.

Beyond identifying the set of events, \textit{Fault Tree Analysis (FTA)} can be applied in conjunction with the causal chain to trace the root causes of hazards. FTA provides a systematic way to decompose a top-level hazardous event into its contributing events, allowing a structured derivation of associated ODD classes and attributes.

\begin{Definition} We define an FTA as a tuple:
\[
FTA = \left \langle te, E, G \right \rangle
\]
where $te \in E$ is the \textit{top event} (typically the hazard), $E$ is a set of events including top, intermediate, and atomic events, and $G$ is a gate function that links a parent event $e_i$ with its cause events $E' \subset E$ through a logical operator gate $opGate \in \{ \wedge, \vee \}$.
\end{Definition}

We outline in Algorithm~\autoref{alg:computeFTA} the recursive procedure to construct an FTA from a given hazard, the set of events derived from HARA, and a causal relation function. 

\begin{algorithm}[!h]
\caption{Compute FTA} \label{alg:computeFTA}
\begin{footnotesize} 
\begin{algorithmic}[1]
\Require $hazard, events, causal\_relation$
\State Define $FTA = \left \langle te, E, G \right \rangle$
\State $te \gets hazard$
\State $events\_stack \gets [te]$

\While{$events\_stack$ is not empty}
    \State $current\_event \gets events\_stack.pop()$
    \State $cause\_events \gets causal\_relation(current\_event)$

    \For{each $event_i \in cause\_events$}
        \If{$event_i \notin E$}
            \State $E.append(event_i)$
        \EndIf

        \If{$event_i$ is not atomic}
            \State $inter\_events \gets causal\_relation(event_i)$
            \State $opGate \gets process\_logical\_gate(event_i, inter\_events)$
            \State $G(event_i, inter\_events, opGate)$
            \State $events\_stack.extend(inter\_events)$
        \ElsIf{$event_i$ is atomic}
            \State Define the set of operational conditions as ODD attributes
        \EndIf
    \EndFor
\EndWhile

\State \Return $FTA$
\end{algorithmic}
\end{footnotesize} 
\end{algorithm}

\begin{example}
    We consider the hazard: \textit{``Danger of collision with pedestrians (hiding objects)"} for the \texttt{AVP} system. The hazard-casual-chain is illustrated in Fig. \ref{fig:risk-based-odd-avp}, while the corresponding FTA is illustrated in Fig. \ref{fig:FTA_AVP}. We also consider the operational conditions as ODD attributes. The identified conditions are presented in Table \ref{tab:ex_Oper_cond_avp}.     

    \begin{figure} [!h]
        \centering
        \includegraphics[width=\linewidth]{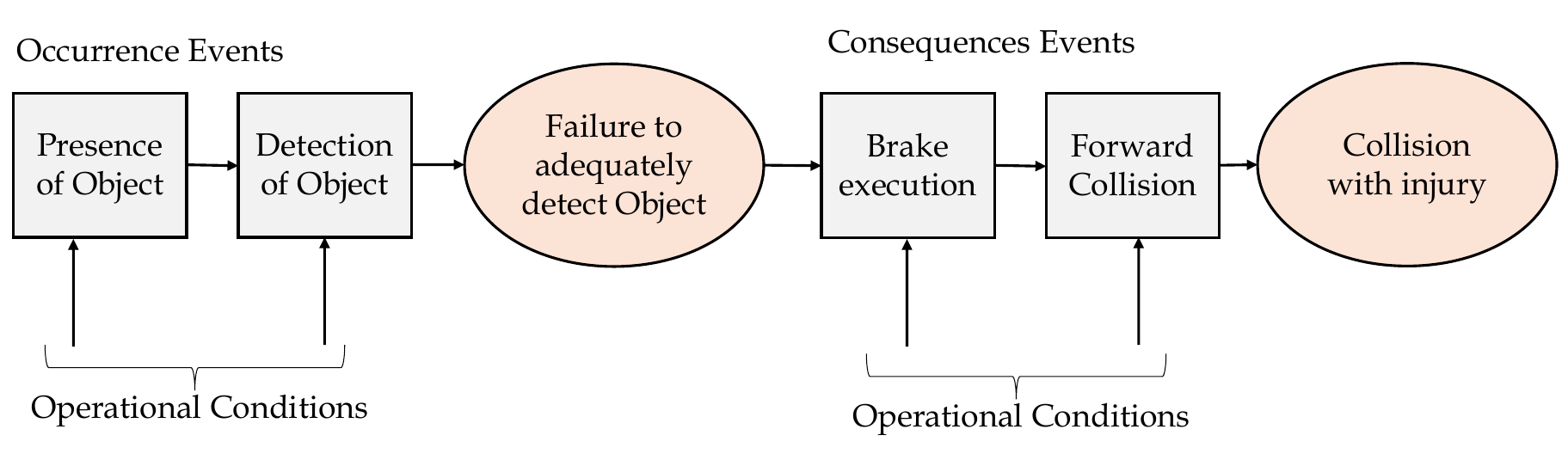}
        \caption{Illustration of a hazard-casual-chain}
        \label{fig:risk-based-odd-avp}
    \end{figure}

    \begin{figure}[!h]
        \centering
        \includegraphics[scale=0.35]{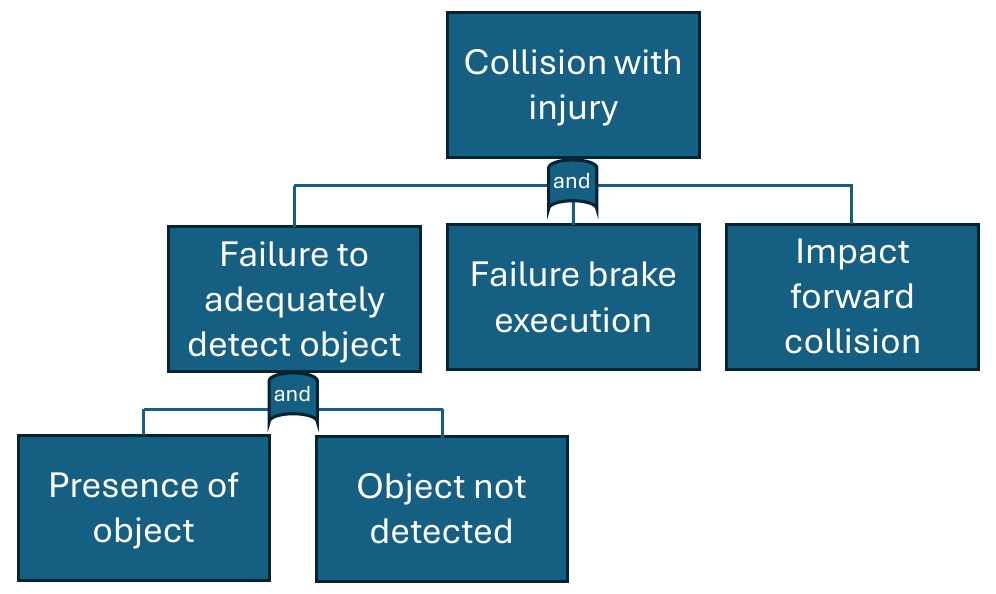}
        \caption{FTA for collision with injury.}
        \label{fig:FTA_AVP}
    \end{figure}
    
    \begin{table}[!h]
        \caption{ODD attributes determination for the \texttt{AVP} system.}
        \label{tab:ex_Oper_cond_avp}
        \scalebox{0.7}{
        \begin{tabular}{|l|l|l|l|}
         \hline
        \multicolumn{1}{|c|}{\textit{\textbf{Hazard}}} & \multicolumn{1}{c|}{\textit{\textbf{Ability}}} & \multicolumn{1}{c|}{\textit{\textbf{Event}}} & \multicolumn{1}{c|}{\textit{\textbf{Oper. Conditions}}} \\ 
        \hline
        \hline
        \multirow{4}{*}{\begin{tabular}[c]{@{}l@{}}Failure to \\ adequately \\ detect \\ object\end{tabular}} & \multirow{4}{*}{\begin{tabular}[c]{@{}l@{}} (A8) Detect \\ obstacle\end{tabular}} & \begin{tabular}[c]{@{}l@{}}Presence of \\ object\end{tabular} & \begin{tabular}[c]{@{}l@{}}Road type = \{Open Parking, Closed Parking\}\\ TrafficParticipant=\{Driver, Pedestrain\}\end{tabular} \\ \cline{3-4} 
         &  & \begin{tabular}[c]{@{}l@{}}Detection of \\ the object\end{tabular} & \begin{tabular}[c]{@{}l@{}}Weather Conditions = \{Rain, Fog, Snow\},\\ Illumination =\{Vehicle\_lighting, Env\_lighting\},\\ Sun\_angle =\{sun\_azimuth\_angle, sun\_altitude\_angle\}\\ Ego\_Speed=\{High, medium, low\}\\ Road Geometry = \{Horizontal\_Curvature, Vertical\_Curvature\}\end{tabular} \\ \cline{3-4} 
         &  & \begin{tabular}[c]{@{}l@{}}Brake \\ execution\end{tabular} & \begin{tabular}[c]{@{}l@{}}Road Surface Conditions= \{Wet, Snow\_covered, Sand\_gravel\}\\ Ego\_Speed=\{High, medium, low\}\end{tabular} \\ \cline{3-4} 
         &  & \begin{tabular}[c]{@{}l@{}}Forward \\ collision\end{tabular} & \begin{tabular}[c]{@{}l@{}}TrafficParticipant=\{Driver, Pedestrain\}\\ Ego\_Speed=\{High, medium, low\}\end{tabular} \\ \hline
        \end{tabular}
        }
    \end{table}

\end{example}

\subsubsection{ODD Attribute Boundaries} Once ODD attributes have been identified through the risk-based approach, it becomes essential to define the precise boundaries within which each attribute is expected to support safe system operation. While the previous step associated specific attributes with their corresponding ODD classes, these attributes must now be further specified in terms of their \textit{data type}, \textit{unit}, and associated \textit{constraints}~\cite{irvine2021two} (cf. Table \ref{tab:oddAttrVal}). The set of constraints define what we refer to as the \textit{ODD attribute boundaries}.

\begin{Definition}
\textit{Let $attr$ be an ODD attribute with a data type $\mathscr{D}$. The function $\mathscr{D}(attr)$ denotes the data type of $attr$. The boundaries of $A$ are formalized as constraints over $\mathscr{D}$, defined as:}
\begin{equation}
attr \bowtie d, \quad \text{where } d \in \mathscr{D} \text{ and } \bowtie \in \{=,<,>,\leq,\geq\}.
\end{equation}
\end{Definition}

At runtime, within the \textit{Operational Domain}, the semantics of an attribute $attr$ are evaluated through an interpretation function $\mathscr{I}$ at a given time $t \in Time$ and location $(x,y) \in Space$, mapping each attribute to a concrete value in its domain:
\begin{align}
\mathscr{I} : attr \times Time \times Space \rightarrow  d,\textbf{ } d\in \mathscr{D}
\end{align}

\begin{example}
    We illustrate in Table \ref{tab:oddAttrVal} the constraints associated to the ODD attributes. These constraints are aligned with the AVSC Best Practice for Describing an Operational Design Domain \cite{automated2020avsc}.

    \begin{table}[!h]
        \caption{ODD attributes with the constraints (domain) for the \texttt{AVP} system.}
        \label{tab:ex_Oper_cond_avp}
        \scalebox{0.835}{
        \begin{tabular}{|l|l|l|l|}
        \hline
        \multicolumn{1}{|c|}{\textit{\textbf{ODD classes}}} & \multicolumn{1}{c|}{\textit{\textbf{Unit}}} & \multicolumn{1}{c|}{\textit{\textbf{Odd attribute}}} & \multicolumn{1}{c|}{\textit{\textbf{Constraint}}} \\ 
        \hline
        \hline
        \multirow{2}{*}{Vehicle\_lighting} & \multirow{2}{*}{Lux} & Vehicle\_lighting\_High & {[}50, 150{]} \\ \cline{3-4} 
         &  & Vehicle\_lighting\_Low & {[}0, 50{[} \\ \hline
        \multirow{5}{*}{Env\_lighting} & \multirow{5}{*}{Lux} & Sunlight & {[}107.527, +{[} \\ \cline{3-4} 
         &  & Very\_Dark\_Day & {[}10.8, 107.527{[} \\ \cline{3-4} 
         &  & Twilight & {[}0.0011, 10.8{[} \\ \cline{3-4} 
         &  & Starlight & {[}0.0001, 0.0011{[} \\ \cline{3-4} 
         &  & Overcast\_Night & {[}0, 0.0001{[} \\ \hline
        \multirow{3}{*}{Rain} & \multirow{3}{*}{Cm per h} & Rain\_light & {[}0, 0.25{[} \\ \cline{3-4} 
         &  & Rain\_Moderate & {[}0.25, 0.77{[} \\ \cline{3-4} 
         &  & Rain\_Heavy & {[}0.77, +{[} \\ \hline
        \multirow{5}{*}{Fog} & \multirow{5}{*}{\begin{tabular}[c]{@{}l@{}}Visibility \\ in meters\end{tabular}} & Fog\_Severity\_1 & {[}1610, +{[} \\ \cline{3-4} 
         &  & Fog\_Severity\_2 & {[}805, 1610{[} \\ \cline{3-4} 
         &  & Fog\_Severity\_3 & {[}244, 805{[} \\ \cline{3-4} 
         &  & Fog\_Severity\_4 & {[}60, 244{[} \\ \cline{3-4} 
         &  & Fog\_Severity\_5 & {[}0, 60{[} \\ \hline
        \multirow{3}{*}{Snow} & \multirow{3}{*}{\begin{tabular}[c]{@{}l@{}}Visibility of light \\ over a distance (Km)\end{tabular}} & Snow\_Light & {[}1, +{[} \\ \cline{3-4} 
         &  & Snow\_Moderate & {[}0.5, 1{[} \\ \cline{3-4} 
         &  & Snow\_Heavy & {[}0, 0.5{[} \\ \hline
        \multirow{3}{*}{Ego\_speed} & \multirow{3}{*}{km/h} & Speed\_High & {[}60, +{[} \\ \cline{3-4} 
         &  & Speed\_Medium & {[}31, 60{]} \\ \cline{3-4} 
         &  & Speed\_Low & {[}0, 31{[} \\ \hline
        \end{tabular}
        }
        \label{tab:oddAttrVal}
    \end{table}
\end{example}

In addition to initial boundary specification, we also consider the  process of \textit{boundary refinement}. This is particularly useful for iteratively improving the definition of safe operating limits based on empirical observations (e.g., simulation traces). Such traces may reveal edge cases where the \texttt{AVP} fails to maintain safety, even if earlier boundaries were satisfied. To support automated refinement, we employ \textit{decision trees} to learn constraints from these operational data points, enabling dynamic updates to the ODD attribute boundaries.

 \begin{example}
     An example of ODD attributes boundaries refined from simulation traces is illustrated in Fig. \ref{fig:refine_ODD}, where  \textit{Yes/No} refers to the conditions with boundaries within the ODD or out of ODD (also referred to as exit-ODD).  

\begin{figure}  
\begin{scriptsize}
\begin{mycode}
###Bounderies: 
IF Vehicle_lighting <= 60.48 AND Fog <= 4828.34 AND Sun_Angle <= 0.19 THEN Yes
IF Vehicle_lighting > 60.48 AND Fog <= 4828.34 AND Sun_Angle <= 0.19 THEN No
IF Vehicle_lighting <= 107.18 AND Fog <= 4828.34 AND Sun_Angle > 0.19 AND Vehicle_speed <= 43.79 THEN Yes
IF Vehicle_lighting <= 107.18 AND Fog <= 4828.34 AND Sun_Angle > 0.19 AND Vehicle_speed > 43.79 THEN No
IF Vehicle_lighting <= 107.18 AND Fog <= 4828.34 AND Sun_Angle > 0.19 AND Vehicle_speed > 56.64 AND Rain <= 1.84 THEN Yes
IF Vehicle_lighting <= 107.18 AND Fog <= 2615.18 AND Sun_Angle > 0.19 AND Vehicle_speed > 56.64 AND Rain > 1.84 THEN Yes
IF Vehicle_lighting <= 107.18 AND Fog > 2615.18 AND Sun_Angle > 0.19 AND Vehicle_speed > 56.64 AND Rain > 1.84 THEN No
...
\end{mycode}
\end{scriptsize}
\caption{Results of the refinement of ODD attributes boundaries.}
\label{fig:refine_ODD}
\end{figure}
\end{example}

\subsection{ODD Formalization}
\label{ODD_Formalization}
The previous sections introduced a methodology for determining relevant ODD classes and attributes by analyzing hazards, causal chains, and FTA. These steps revealed a rich structure of operational conditions and measurable attributes that directly influence the safety of the system. To enable formal reasoning over these conditions, and to support automated consistency checks, querying, and traceability across system components, we now define a formal representation of the ODD. 

In this section, we formalize the ODD using an ontological approach grounded in \textit{DL} and aligned with the \textit{OpenODD} ontology. The identified ODD classes (e.g., \texttt{Rain}, \texttt{Vehicle\_lighting}, \texttt{Ego\_speed}) are modeled as \textit{OddClass}, while their measurable characteristics (e.g., \texttt{Vehicle\_lighting\_High}, \texttt{Fog\_Severity\_1}) are modeled as \textit{OddAttributes}. This structured formalization allows precise, machine-interpretable semantics and supports the integration of ODD knowledge into safety assurance pipelines. The following axioms establish the fundamental relationships between ODD classes:

{\small
\begin{flalign*}
(A_1) & \quad \top \sqsubseteq \forall subClassOf.(OddClass) & \\
(A_2) & \quad \exists subClassOf \top \sqsubseteq OddClass &
\end{flalign*} 
}
Axioms ($A_1$) and ($A_2$) state that every individual in the domain $\top$ is such that if it is related to another individual via the $subClassOf$ property, the related individual must be an $OddClass$. 

\begin{example} \textit{To illustrate the application of these axioms, consider the example of \texttt{weather\_conditions}. The structure of ODD classes for weather conditions can be defined as follows:}  ``\texttt{ODD} $\rightarrow$ \texttt{Environment\_conditions} $\rightarrow$ \texttt{Weather\_conditions} $\rightarrow$ $ $\texttt{\{Rain, Fog, Snow\}}".  \\

\textit{This hierarchy is represented in the ABox as:}
{\small
\begin{flalign*}
& \quad (Weather\_conditions, Environment\_conditions): subClassOf & \\
& \quad (Rain, Weather\_conditions): subClassOf & \\
& \quad (Fog, Weather\_conditions): subClassOf & \\
& \quad (Snow, Weather\_conditions): subClassOf & 
\end{flalign*}
}
\textit{This definition ensures that \texttt{Rain, Fog}, and \textit{Snow} are subclasses of \texttt{Weather\_conditions}, which itself is a subclass of \texttt{Environment\_conditions}.}
\end{example}

Further formalization of the ODD involves defining \textit{attributes} and their respective domains. The following axioms define these relations:

{\small
\begin{flalign*}
(A_3) & \quad \top \sqsubseteq \forall hasAttribute.(OddAttribute) & \\
(A_4) & \quad \exists hasAttribute \top \sqsubseteq OddClass &\\
(A_5) & \quad \top \sqsubseteq \forall hasDomain.(Unit \sqcup Constraint) & \\
(A_6) & \quad \exists hasDomain \top \sqsubseteq OddAttribute &
\end{flalign*}
}

Axioms $(A_3)$ and $(A_4)$ ensure that each ODD class has an associated attribute, while axioms $(A_5)$ and $(A_6)$ define that attributes have a domain consisting of both a unit and a constraint.

\begin{example} \textit{Defining a Rain Attribute, for instance, the ODD attribute \texttt{Rain\_heavy} can be formally described in the ABox as follows:}

{\small
\begin{flalign*}
& \quad (Rain, Rain\_heavy): hasAttribute & \\
& \quad (Rain\_heavy, ``cm/h"): hasDomain & \\
& \quad (Rain\_heavy, ``\geq0.77"): hasDomain & \\
& \quad ``cm/h": Unit \\
& \quad ``>=0.77": Constraint & 
\end{flalign*}
}
\textit{This representation indicates that the \texttt{Rain} class has an attribute \texttt{Rain\_heavy}, which is measured in \texttt{cm/h} and constrained to values \texttt{$\geq$0.77cm/h}.}
\end{example}

To integrate risk analysis into the ODD formalization, we extend the initial set of axioms to capture the semantics of a \textit{hazard-cause-effect} framework. Specifically, axioms $(A_{7})$ to $(A_{23})$ formalize the relationships between ODD attributes and various event types (occurrence, hazardous, and consequence events).

{\small
\begin{flalign*}
(A_7) & \quad \top \sqsubseteq \forall hasOperCond.(OddAttribute ) & \\
(A_8) & \quad \exists hasOperCond \top \sqsubseteq Event \sqcup OccurranceEvent \sqcup &\\
& ConsequenceEvent & 
\end{flalign*}
}
The relation $hasOperCond$ associates events with ODD attributes.

{\small
\begin{flalign*}
(A_9) & \quad \top \sqsubseteq \forall hasOccurrenceEvent.(OccurrenceEvent) & \\
(A_{10}) & \quad \exists hasOccuranceEvent \top \sqsubseteq Event  &
\end{flalign*}
}
The relation $hasOccurrenceEvent$ defines the link between general events and occurrence events.

{\small
\begin{flalign*}
(A_{11}) & \quad \top \sqsubseteq \forall hasConsequenceEvent.(ConsequenceEvent) & \\
(A_{12}) & \quad \exists hasConsequenceEvent \top \sqsubseteq Event  &
\end{flalign*}
}
The relation $hasConsequenceEvent$ establishes the connection between general events and consequence events.

{\small
\begin{flalign*}
(A_{13}) & \quad \top \sqsubseteq \forall trigger.(HazardousEvent) & \\
(A_{14}) & \quad \exists trigger \top \sqsubseteq OccuranceEvent  &
\end{flalign*}
}
The relation $trigger$ represents the causal link between an occurrence event and a hazardous event.
The following axioms establish dependency relationships among occurrence, hazardous, and consequence events.

{\small
\begin{flalign*}
(A_{15}) & \quad T \sqsubseteq \forall dependsOnOccurrence.(OccurrenceEvents) & \\
(A_{16}) & \quad \exists dependsOnOccurrence \top \sqsubseteq (OccurrenceEvents \sqcup & \\
& HazardousEvents) &\\
(A_{17}) & \quad T \sqsubseteq \forall dependsOnHazardous.(HazardousEvents) & \\
(A_{18}) & \quad \exists dependsOnHazardous \top \sqsubseteq (HazardousEvents \sqcup & \\
& ConsequenceEvents)  &\\
(A_{19}) & \quad T \sqsubseteq \forall dependsOnConsequence.(ConsequenceEvents) & \\
(A_{20}) & \quad \exists dependsOnConsequence \top \sqsubseteq (ConsequenceEvents)  &
\end{flalign*}
}
The final axioms classify events based on their dependencies.

{\small
\begin{flalign*}
(A_{21}) & \quad OccurrenceEvent \equiv Event \sqcap (\neg dependsOnHazardous \sqcup  &\\
& \neg dependsOnConsequence) &\\
(A_{22}) & \quad ConsequenceEvent \equiv Event \sqcap \neg dependsOnOccurrence  &\\
(A_{23}) & \quad HazardousEvent \equiv  Event \sqcap \neg dependsOnConsequence &
\end{flalign*}
}

\section{Identification and formalization of Assurance Goals}
\label{subsec:safety-assurance}
In \cite{wozniak2020safety}, the authors propose the use of safety case patterns to structure the safety argumentation for DNN components. These patterns serve as reusable templates that guide the construction of arguments concerning the correctness and dependability of DNN component implementations. Similarly, the AMLAS (Assurance of Machine Learning in Autonomous Systems) framework \cite{hawkins2021guidance} introduces a systematic process and a set of safety case patterns for: (1) integrating safety assurance throughout the development lifecycle of ML components, and (2) generating structured evidence to support claims about their safe integration into autonomous systems.

\subsubsection{Identification of Assurance Goals}
\begin{figure}[!h]
    \centering
    \includegraphics[scale=0.5]{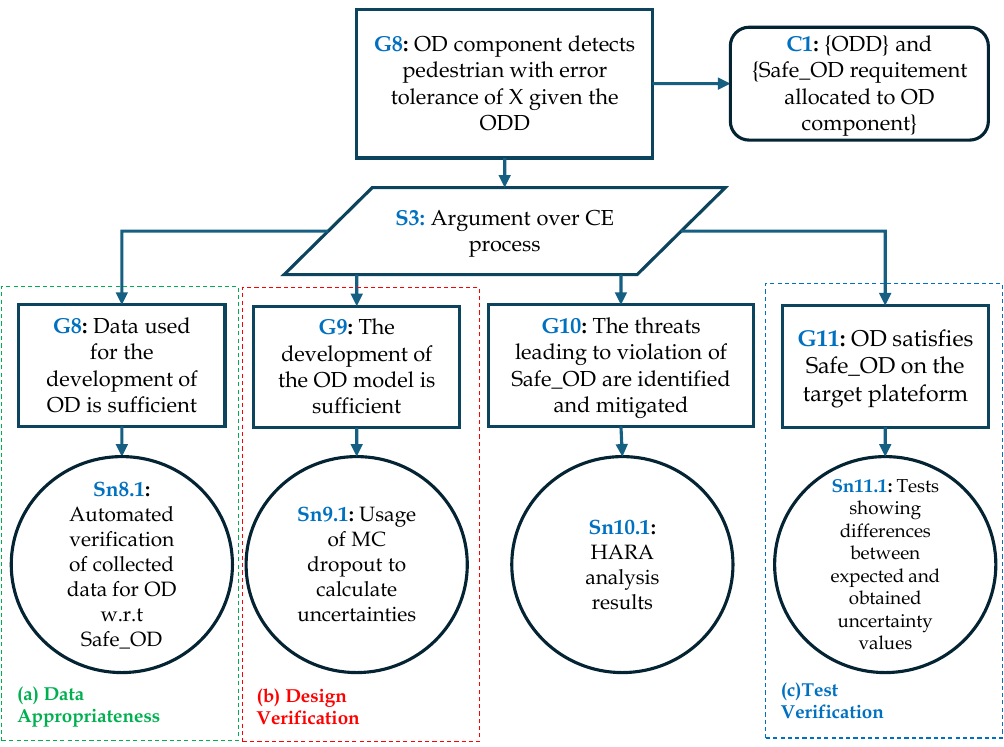}
    \caption{Top-level of GSN-based pattern for safety case construction for \texttt{OD} component of the \texttt{AVP} system \cite{sorokin2024towards}}
    \label{fig:ml-verif-gsn}
\end{figure}

For the \texttt{AVP} system, the approach presented in \cite{sorokin2024towards} adopts these established safety case patterns to construct a structured safety argument. This enables a systematic demonstration that the DNN within the \texttt{OD} component within acceptable safety boundaries (see Fig.~\ref{fig:ml-verif-gsn}). Specifically, the GSN-based patterns are used to represent the \emph{decomposition strategy} applied throughout the continuous engineering process of the \texttt{OD} component. This includes assurance arguments related to \emph{(a) data appropriateness}, the \emph{(b) design and training} of the DNN, and \emph{(c) test verification}. 

For data appropriateness, it is essential to ensure that the collected datasets satisfy the functional requirements of the DNN. This includes verifying that the data is properly labeled, representative of the target domain, and provides sufficient coverage across training, validation, and testing phases. Additionally, the quantity of data samples must be adequate to support robust model development and evaluation. Regarding the design of the DNN component, we must verify its correctness, verifiability, and robustness with respect to the intended model architecture. 

In this paper, we focus on the solutions \texttt{Sn8.1, Sn9.1, Sn11.1} within the argument pattern shown in Fig. \ref{fig:ml-verif-gsn}. The objective is to construct a confident argument corresponding to these solutions, as illustrated in Section~\ref{sec:Confidence_Model}.

\subsubsection{Formalization of Assurance Goals}
\label{sec:formalization_sg}
This section introduces the formalization of the safety argumentation using an ontological approach. The resulting formalization serves as the foundation for computing confidence values over the safety case and linking them to the underlying operational and model-level evidence.  Axioms $(A_{24})-(A_{25})$ relate the hazardous behaviours identified in Section \ref{ODD_Formalization} with a top level goal. Axioms $(A_{26})-(A_{29})$ represent the GSN goals, strategies and solutions, together with the $supportedBy$ relation.

{\small
\begin{flalign*}
(A_{24}) & \quad \top \sqsubseteq \forall relatedTo.(HazardousEvent) & \\
(A_{25}) & \quad \exists relatedTo  \top \sqsubseteq TopLevelGoal &\\
(A_{26}) & \quad \top \sqsubseteq \forall supportedBy.(Goal \sqcup Strategy \sqcup Solution) & \\
(A_{27}) & \quad \exists supportedBy \top \sqsubseteq Goal \sqcup Strategy &\\
(A_{28}) & \quad  supportedBy^- \equiv supports &\\
(A_{29}) & \quad supportedBy \sqsubseteq supportedBy &
\end{flalign*}
}
Axioms $(A_{30})-(A_{32})$ state the inference relation between goals.

{\small
\begin{flalign*}
(A_{30}) & \quad \top \sqsubseteq \forall hasInference.Goal & \\
(A_{31}) & \quad \exists hasInference \top \sqsubseteq Goal  &\\
(A_{32}) & \quad hasInference \sqsubseteq supportedBy &
\end{flalign*}
}
Axioms $(A_{33})-(A_{35})$ state the evidence relation between goals.

{\small
\begin{flalign*}
(A_{33}) & \quad \top \sqsubseteq \forall hasEvidence.Evidence & \\
(A_{34}) & \quad \exists hasEvidence \top \sqsubseteq Goal  &\\
(A_{35}) & \quad hasEvidence \sqsubseteq supportedBy &
\end{flalign*}
}
Axioms $(A_{36})-(A_{37})$ associates the confidence to the objective node.

{\small
\begin{flalign*}
(A_{36}) & \quad \top \sqsubseteq \forall hasConfidence.ObjNode & \\
(A_{37}) & \quad \exists hasConfidence \top \sqsubseteq (Goal \sqcap Solution)  &
\end{flalign*}
}
Axioms $(A_{38})-(A_{39})$ define the support and top level goals.

{\small
\begin{flalign*}
(A_{38}) & \quad SupportGoal \equiv Goal \sqcap \exists supports.\top & \\
(A_{39}) & \quad TopLevelGoal \equiv Goal \sqcap (\neg SupportGoal)  &
\end{flalign*}
}
Axioms $(A_{40}) - (A_{41})$ define the statement with a text. 

{\small
\begin{flalign*}
(A_{40}) & \quad \top \sqsubseteq \forall hasText.String& \\
(A_{41}) & \quad \exists hasText.Statement \sqsubseteq \top  &
\end{flalign*}
}

\section{Confidence Model Creation}
\label{sec:Confidence_Model}
This section presents the methodology for creating the confidence model associated with the assurance solutions discussed in Section~\ref{subsec:safety-assurance}. In \cite{hawkins2011new}, the authors refer to this as \textit{Assurance Claim Points (ACP)} with the type \textit{Asserted solution}. In this paper, we refer to the ACP as \textit{confidence model}. 

\begin{Definition}
An ACP identifies a specific location in the safety argument, typically a link to a solution element, where a confidence argument is attached to justify that the supporting evidence is sufficient and trustworthy for the associated claim.
\end{Definition}

To quantify the confidence associated to the solutions \texttt{Sn8.1, Sn9.1, Sn11.1} within the argument pattern shown in Fig. \ref{fig:ml-verif-gsn}, we leverage a probabilistic reasoning framework based on BNs. 

The proposed confidence model is constructed by systematically integrating the outcomes of the risk-based hazard analysis from Section~\ref{subsubsec:risk-based-determination} with the structured safety argumentation patterns introduced in Section~\ref{subsec:safety-assurance}. Specifically, we define a BN in which nodes represent relevant assurance-related variables (e.g., data appropriateness, model robustness, test coverage), while edges encode the causal and logical dependencies derived from FTA and argument patterns.

\begin{Definition}
Bayesian Networks (BNs) are graphical models that represent probabilistic relationships among a set of random variables in an uncertain domain~\cite{shih2018formal}. In a BN, each node corresponds to a random variable, and each edge encodes a conditional dependency between the connected variables. Formally, a BN is defined as a directed acyclic graph (DAG) annotated with a set of conditional probability distributions. Specifically, a BN is a tuple $B = \langle G, \Theta \rangle$, where $G$ is a DAG whose nodes $x_1, x_2, \dots, x_n$ represent random variables, and the edges indicate direct probabilistic dependencies among them. The set $\Theta$ contains the Conditional Probability Tables (CPTs) that quantify the effects of the dependencies~\cite{shih2018formal}.
\end{Definition}

\subsubsection{Constructing the Initial BN}
A key question that arises is: \textit{``How can we construct the initial BN to quantify the confidence in the assurance solutions?"} To address this, we build upon the causality analysis presented in Section~\ref{subsubsec:risk-based-determination}, complemented by the FTA. Together, these techniques enable the systematic identification of causal dependencies and failure propagation paths, which serve as the foundation for the structure of the initial BN.

As described in Section~\ref{subsubsec:risk-based-determination}, we generate a fault tree structure defined as $FTA = \langle te, E, G \rangle$, where $te$ is the top event (i.e., the hazard), $E$ is the set of all events, and $G$ represents the logical gates connecting these events. The mapping from FTA to a BN is defined as follows:

\begin{itemize}
    \item \textit{Events ($e \in E$):} Each event $e \in E$ is mapped to a corresponding node $X$ in the BN. The top event $te$ is modeled as the objective node, typically represented as the terminal (bottom-most) node in the BN structure.
    
    \item \textit{Causal Dependencies:} The causal relationships represented by edges in the FTA are translated into directed edges in the BN, capturing conditional dependencies. FTA logic gates are encoded in the BN using Conditional Probability Tables (CPTs):
    \begin{itemize}
        \item \textit{\textbf{$\wedge$-Gate (AND):}} The probability of the output event occurring is computed as the product of the probabilities of its input events.
        \item \textit{\textbf{$\vee$-Gate (OR):}} The probability of the output event occurring is calculated based on the assumption that at least one of the input events occurs.
    \end{itemize}
\end{itemize}

The initial CPTs are typically defined by domain experts using insights from the FTA and risk analysis. These tables can be refined and updated through empirical observations or simulation data, enabling more accurate instantiation and fine-tuning. As a central component of the BN, CPTs govern the behavior of each node by quantifying how its state depends on the states of its parent nodes, thus enabling robust probabilistic inference across the network. We will illustrate in the following sections the initial confidence model related to the solutions \texttt{Sn8.1, Sn9.1} and \texttt{Sn11.1} (cf. Section \ref{subsec:safety-assurance}). 
\begin{itemize}
    \item \textbf{\textit{Data appropriateness.}} To evaluate the relevance and quality of training data with respect to the safety goals (cf. Fig. \ref{fig:ml-verif-gsn}(a)), we assess its \emph{completeness}. Completeness, in this context, refers to the extent to which the collected dataset covers scenarios pertinent to the stated safety goals. A dataset is considered complete if it includes representative instances of these scenarios. This requires (i) identifying the relevant scenarios for the safety objective, and (ii) applying a metric that quantifies how well the dataset covers them. For this purpose, we use the \emph{scenario coverage metric} introduced in~\cite{cheng2018quantitative}, which is calculated as:
        \begin{equation}
            m = \frac{{n_{occurrences}}}{n_{total}}
        \end{equation}
        Here, $m$ denotes the scenario coverage metric, where $n_{occurrences}$ represents the number of times relevant scenarios appear in the dataset, and $n_{total}$ is the total number of data samples.        
        Once the relevant scenarios are identified, we use the BN shown in Fig. \ref{fig:goodData} to compute the completeness of the dataset with respect to those scenarios.
        
        \begin{figure}[!h]
            \centering
            \vspace{0.5cm} 
            \scalebox{0.5}{
                \begin{tikzpicture}[
                    node distance=2.5cm, 
                    bend angle=50, 
                    every node/.style={scale=1},
                    place/.style={circle, draw=black, thick, minimum size=20mm, fill=gray!10, line width=1pt},
                    transition/.style={rectangle, draw=black, thick, minimum width=10mm, minimum height=4mm, fill=black!20}, 
                    >=stealth,
                    line width=2.5pt
                ]
                \node[place] (Feat_1) {$Feat_1$};
                \node[place] (Feat_2) [right of=Feat_1, xshift=1cm] {$Feat_2$};
                \node[place] (Feat_i) [below of=Feat_1, xshift=2cm] {$Feat_i$};
                \node[place] (ObjFun) [below of=Feat_i, yshift=-0.5cm] {$ObjFun$};
                \node[place] (OddFC) [right of=ObjFun, xshift=1cm] {$OddFC$};
                \node[place] (OddSuff) [below of=ObjFun, xshift=2cm, yshift=-0.5cm] {$OddSuff$};
                \node[place] (DataMetric) [right of=OddSuff, xshift=2cm] {$DataMetric$};
                \node[place, fill=blue!20] (DataComp) [below of=OddSuff, yshift=-0.5cm, xshift=1.3cm] {$DataComp$};
        
                \draw[->] (Feat_1) -- (Feat_i);
                \draw[->] (Feat_2) -- (Feat_i);
                \draw[->] (Feat_i) -- (ObjFun);
                \draw[->] (ObjFun) -- (OddSuff);
                \draw[->] (OddFC) -- (OddSuff);
                \draw[->] (OddSuff) -- (DataComp);
                \draw[->] (DataMetric) -- (DataComp);
                \end{tikzpicture}
            }    
            \caption{BN for data appropriateness quantification.}
            \label{fig:goodData}
        \end{figure}
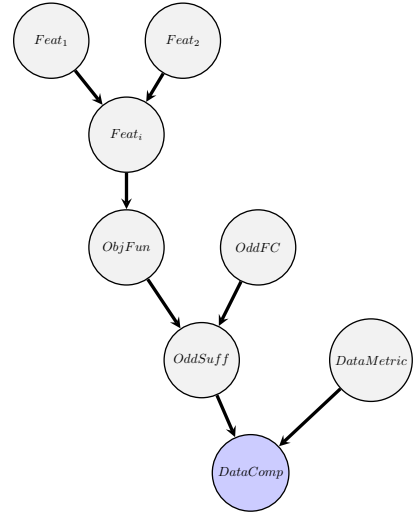
        The CPT corresponding to the \texttt{DataMetric} node is derived using the scenario coverage metric $m$, while the CPTs for the \texttt{OddFC}, \texttt{OddSuff}  and \texttt{DataComp} nodes are initially defined using expert knowledge. These CPTs can be further refined using operational data. While this BN is scoped for data completeness, it can be extended using the same approach to cover data accuracy, data relevance and data balance \cite{hawkins2021guidance}.

    \item \textbf{\textit{Model robustness.}} Building on the data appropriateness and ODD analysis discussed earlier, we extend the BN to include the model's predictive uncertainty. Specifically, we use \emph{Monte Carlo Dropout}  \cite{gal2016dropout} to estimate uncertainty in DNNs, as it enables stochastic predictions at test time by activating dropout layers, normally disabled during inference. This provides a distribution over outputs, capturing variability due to model uncertainty. As shown in Figure~\ref{fig:model-monte}, the \texttt{BnModelUnc} node reflects this uncertainty and connects with other factors (e.g., data completeness and ODD sufficiency).
    
    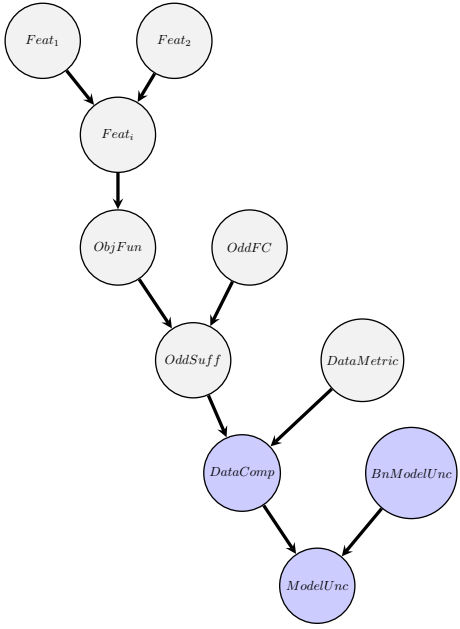
\begin{figure}[!h]
        \centering
        \vspace{0.5cm} 
        \scalebox{0.5}{ 
            \begin{tikzpicture}[
                node distance=2.5cm, 
                bend angle=50, 
                every node/.style={scale=1},
                place/.style={circle, draw=black, thick, minimum size=20mm, fill=gray!10, line width=1pt}, 
                transition/.style={rectangle, draw=black, thick, minimum width=10mm, minimum height=4mm, fill=black!20}, 
                >=stealth,
                line width=2.5pt
            ]
            \node[place] (Feat_1) {$Feat_1$};
            \node[place] (Feat_2) [right of=Feat_1, xshift=1cm] {$Feat_2$};
            \node[place] (Feat_i) [below of=Feat_1, xshift=2cm] {$Feat_i$};
            \node[place] (ObjFun) [below of=Feat_i, yshift=-0.5cm] {$ObjFun$};
            \node[place] (OddFC) [right of=ObjFun, xshift=1cm] {$OddFC$};
            \node[place] (OddSuff) [below of=ObjFun, xshift=2cm, yshift=-0.5cm] {$OddSuff$};
            \node[place] (DataMetric) [right of=OddSuff, xshift=2cm] {$DataMetric$};
            \node[place] (DataComp) [below of=OddSuff, yshift=-0.5cm, xshift=1.3cm, fill=blue!20] {$DataComp$};
            \node[place] (BnModelUnc) [right of=DataComp, xshift=2cm, fill=blue!20] {$BnModelUnc$};
            \node[place] (ModelUnc) [below of=DataComp, yshift=-0.5cm, xshift=2cm, fill=blue!20] {$ModelUnc$};

            \draw[->] (Feat_1) -- (Feat_i) node[midway, above] {};
            \draw[->] (Feat_2) -- (Feat_i) node[midway, above] {};
            \draw[->] (Feat_i) -- (ObjFun) node[midway, above] {};
            \draw[->] (ObjFun) -- (OddSuff) node[midway, above] {};
            \draw[->] (OddFC) -- (OddSuff) node[midway, above] {};
            \draw[->] (OddSuff) -- (DataComp) node[midway, above] {};
            \draw[->] (DataMetric) -- (DataComp) node[midway, above] {};
            \draw[->] (DataComp) -- (ModelUnc) node[midway, above] {};
            \draw[->] (BnModelUnc) -- (ModelUnc) node[midway, above] {};
            
            \end{tikzpicture}
        }    
       \caption{BN for model robustness quantification.}
        \label{fig:model-monte}
    \end{figure}

    \item \textbf{\textit{Testing adequacy.}} Following the evaluation of data appropriateness and model uncertainty, we now consider the verification of the implemented DNN model, particularly to ensure that its behavior aligns with design expectations and performs reliably on the target platform. 
    This requires test cases that validate both functional correctness and runtime performance. A key aspect is comparing model outputs against expected results (i.e., ground truth) using distance metrics such as \emph{Jaccard}, \emph{Hamming}, \emph{Manhattan (L1)}, or \emph{Euclidean distance}, depending on the nature of the data.  As in earlier sections, we distinguish between \textit{training} and \textit{testing} datasets, each requiring a separate evaluation of data appropriateness. Although the approach is structurally the same, the resulting CPTs will differ due to the underlying data variations.
    
    Fig.~\ref{fig:training-unc} presents the extended BN used to compute the test-related uncertainty. This network incorporates the output of model uncertainty (\emph{BNModelUnc}) and testing adequacy to refine the overall confidence in the ML component's implementation.

    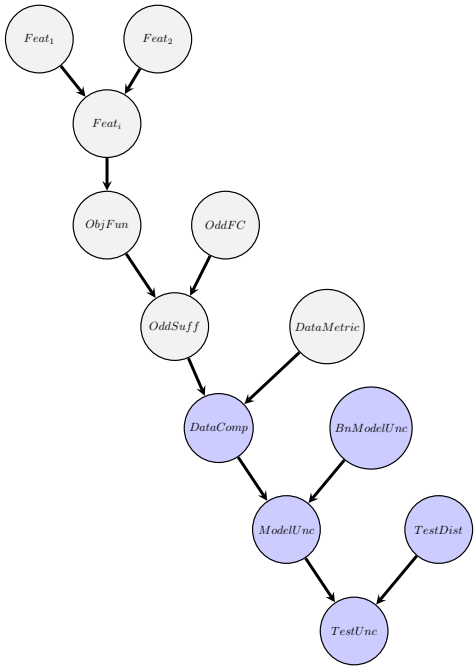
\begin{figure}[!h]
        \centering
        \vspace{0.5cm} 
        \scalebox{0.45}{ 
            \begin{tikzpicture}[
                node distance=2.5cm, 
                bend angle=50, 
                every node/.style={scale=1},
                place/.style={circle, draw=black, thick, minimum size=20mm, fill=gray!10, line width=1pt}, 
                transition/.style={rectangle, draw=black, thick, minimum width=10mm, minimum height=4mm, fill=black!20}, 
                >=stealth,
                line width=2.5pt
            ]
            \node[place] (Feat_1) {$Feat_1$};
            \node[place] (Feat_2) [right of=Feat_1, xshift=1cm] {$Feat_2$};
            \node[place] (Feat_i) [below of=Feat_1, xshift=2cm] {$Feat_i$};
            \node[place] (ObjFun) [below of=Feat_i, yshift=-0.5cm] {$ObjFun$};
            \node[place] (OddFC) [right of=ObjFun, xshift=1cm] {$OddFC$};
            \node[place] (OddSuff) [below of=ObjFun, xshift=2cm, yshift=-0.5cm] {$OddSuff$};
            \node[place] (DataMetric) [right of=OddSuff, xshift=2cm] {$DataMetric$};
            \node[place] (DataComp) [below of=OddSuff, yshift=-0.5cm, xshift=1.3cm, fill=blue!20] {$DataComp$};
            \node[place] (BnModelUnc) [right of=DataComp, xshift=2cm, fill=blue!20] {$BnModelUnc$};
            \node[place] (ModelUnc) [below of=DataComp, yshift=-0.5cm, xshift=2cm, fill=blue!20] {$ModelUnc$};

            \node[place] (TestDist) [right of=ModelUnc, xshift=2cm, fill=blue!20] {$TestDist$};
            
            \node[place] (TestUnc) [below of=ModelUnc, yshift=-0.5cm, xshift=2cm, fill=blue!20] {$TestUnc$};

            \draw[->] (Feat_1) -- (Feat_i) node[midway, above] {};
            \draw[->] (Feat_2) -- (Feat_i) node[midway, above] {};
            \draw[->] (Feat_i) -- (ObjFun) node[midway, above] {};
            \draw[->] (ObjFun) -- (OddSuff) node[midway, above] {};
            \draw[->] (OddFC) -- (OddSuff) node[midway, above] {};
            \draw[->] (OddSuff) -- (DataComp) node[midway, above] {};
            \draw[->] (DataMetric) -- (DataComp) node[midway, above] {};
            \draw[->] (DataComp) -- (ModelUnc) node[midway, above] {};
            \draw[->] (BnModelUnc) -- (ModelUnc) node[midway, above] {};
            \draw[->] (TestDist) -- (TestUnc) node[midway, above] {};
            \draw[->] (ModelUnc) -- (TestUnc) node[midway, above] {};

            \end{tikzpicture}
        }    
        \caption{BN to capture the testing adequacy.}
        \label{fig:training-unc}
    \end{figure}

\end{itemize}

\subsubsection{Formalization of Confidence Model} To enable integration of the confidence model with previously formalized ODD and safety argumentation structures, we now provide the DL axiomatization of the initial BN.  This formalization complements the BN models introduced in the previous sections by capturing the semantic relationships between uncertainty sources (e.g., data, model, and environment) and their influence on assurance goals. Specifically, each node in the BN whether representing data completeness, model uncertainty, or ODD compliance, is modeled as an individual concept, with directed dependencies encoded via DL relations. The core axioms used to represent the structure of the BN are as follows:  

{\small
\begin{flalign*}
(A_{42}) & \quad \top \sqsubseteq\forall dependsOn.(Node)  & \\
(A_{43}) & \quad \exists dependsOn \top \sqsubseteq Node  & 
\end{flalign*}
}

The relation $dependsOn$ represents the dependency between nodes in the BN.
{\small
\begin{flalign*}
(A_{44}) & \quad \top \sqsubseteq \forall hasCPT.(CptTable) & \\
(A_{45}) & \quad \exists hasCPT \top \sqsubseteq Node  & 
\end{flalign*}
}

The relation $hasCPT$ relates a conditional probability table to a node.
{\small
\begin{flalign*}
(A_{46}) & \quad \top \sqsubseteq \forall hasACP.(Value) & \\
(A_{47}) & \quad \exists hasACP \top \sqsubseteq ObjNode & \\
(A_{48}) & \quad ObjNode \equiv Node \sqcap (dependsOn^{-}) & 
\end{flalign*}
}

The relation $ObjNode$ defines what is the objective node representing the objective function in the BN. The objective node relates with the GSN axiomatization defined in Section \ref{sec:formalization_sg}.

\begin{example}
    The corresponding initial BN derived from this FTA and the initial ODD is illustrated in Fig. \ref{fig:Intial_BN}.
    \begin{figure}[!h]
    \hspace*{-1cm}
        \centering
        \includegraphics[scale=0.3]{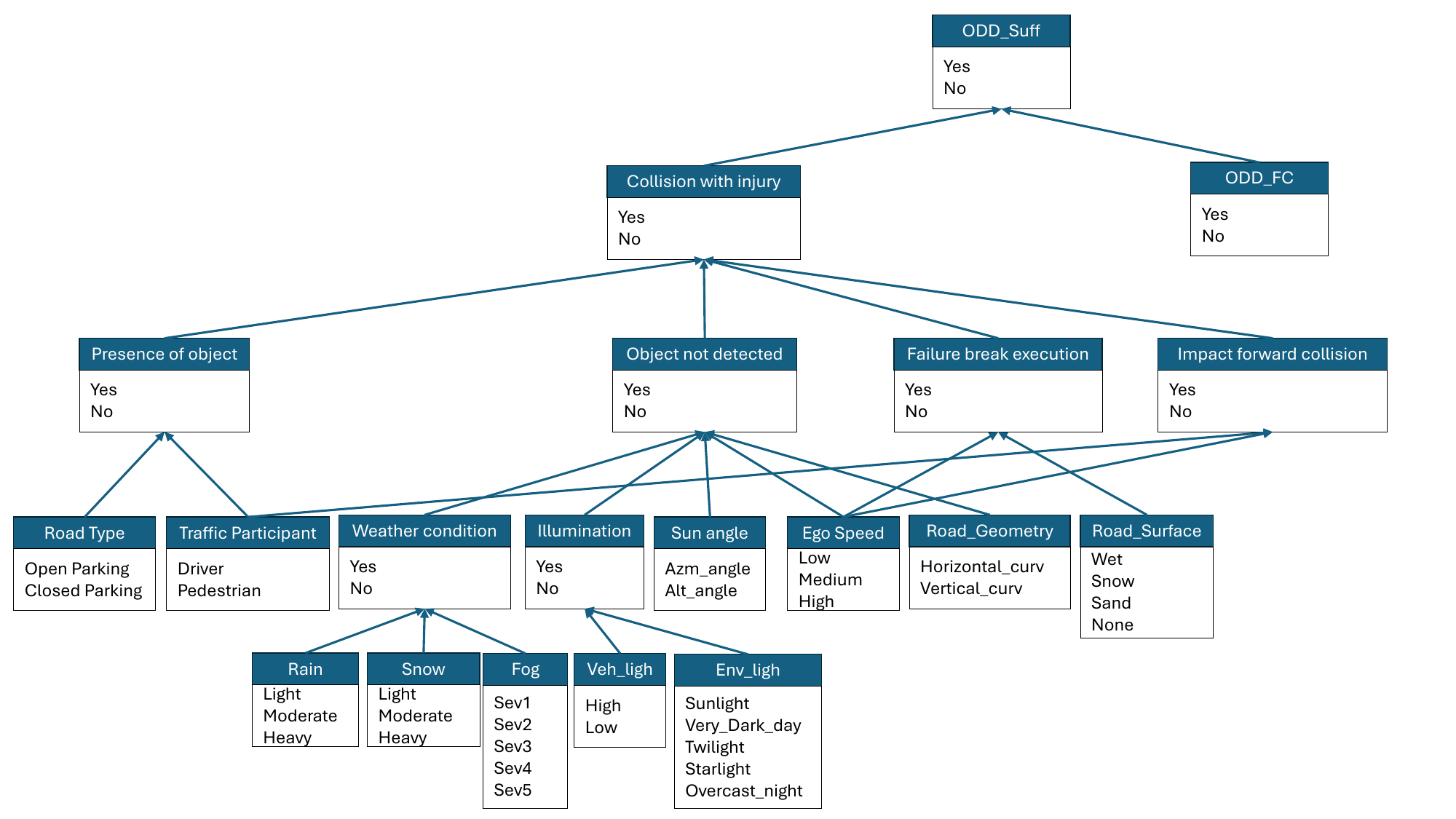}
        \caption{Initial BN related to \texttt{ODD\_Suff}.}
        \label{fig:Intial_BN}
    \end{figure}
\end{example}

\section{Computing the Confidence during Runtime} 
\label{subsec:operation-time-confidence}

To support runtime assurance of the \texttt{AVP} system, we leverage the BN models developed during design-time, covering data appropriateness, model uncertainty, and testing adequacy, to compute confidence values dynamically during system operation. This is achieved by evaluating the \emph{posterior probability} over the BN using the current environmental and operational context, as defined by the instantiated ODD attributes. Formally, given a BN with $n$ random variables $\{x_1, x_2, \dots, x_n\}$, where each node $x_i$ has a set of parent nodes $Pa(x_i)$, the joint probability distribution is defined as:
\begin{equation} 
    \mathbb{P}(x_1, x_2, \dots, x_n) = \prod _{i=1}^n \mathbb{P}(X_i \mid Pa(X_i)) 
\end{equation}
The posterior probability for a query variable $q$ given evidence $e$ (e.g., real-time sensor readings or context updates) is computed using Bayes’ theorem:

\begin{equation} 
    \mathbb{P}(q \mid e) = \frac{\mathbb{P}(q,e)}{\mathbb{P}(e)} 
\end{equation}
Here, $\mathbb{P}(q,e)$ is the joint probability of the query and evidence, and $\mathbb{P}(e)$ is the marginal probability of the evidence. At runtime, we instantiate the Bayesian models defined in Fig. \ref{fig:goodData}, \ref{fig:model-monte}, and \ref{fig:training-unc}, using current ODD observations, such as weather, lighting, road conditions, or ego vehicle speed, as evidence. For example, if the ODD class \texttt{weather = \{sunny, cloudy, rainy\}} and the system detects rain at time $t_i$, then \texttt{weather = rainy} becomes part of the evidence set.

The posterior probabilities computed from these BNs provide a confidence score over each assurance dimension (data, model, testing), which can be used to assess the system’s current level of trustworthiness. These values feed into the overall safety case argumentation framework described in Section~\ref{subsec:safety-assurance}, enabling a continuous, context-aware validation of the system’s safe operation.

\section{Implementation}
\label{sec:Implementation}

To validate the proposed methodology, we developed a proof-of-concept implementation, \texttt{Open-DSC}\footnote{\url{https://git.fortiss.org/depai/dyn-sc}}, which encompasses both the design-time modeling and runtime evaluation aspects of dynamic safety assurance.

\subsection{Formalization and Reasoning}

For the modeling and reasoning over the \textit{Safety Ontology}, we employed \texttt{Protégé}\footnote{\url{https://protege.stanford.edu/}} to construct and manage ontological representations of the ODD,  safety arguments, and confidence models. The ontology is grounded in DL and aligned with the OpenODD standard. The taxonomy of safety-related classes and their interrelations is illustrated in Fig.~\ref{fig:taxonomy-classes} and Fig. .

\begin{figure}[!h]
    \centering
    \includegraphics[width=0.8\linewidth]{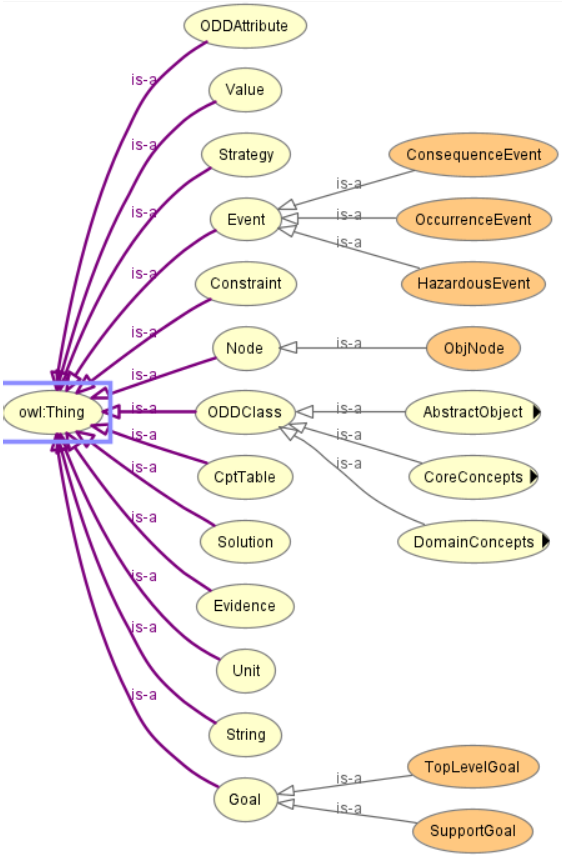}
    \caption{Taxonomy of classes within the safety ontology, formalizing the relationships among ODD, safety arguments, and confidence models.}
    \label{fig:taxonomy-classes}
\end{figure}

\begin{figure}[!h]
    \centering
    \includegraphics[width=0.55\linewidth]{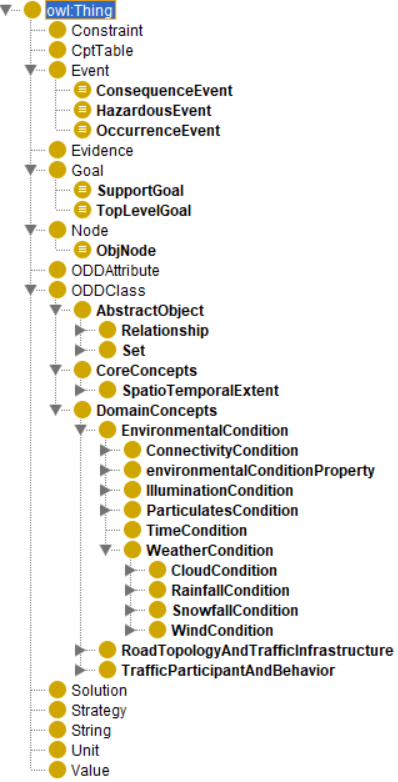}
    \caption{Taxonomy of classes within the safety ontology in \texttt{Protégé}.}
    \label{fig:taxonomy-classes}
\end{figure}

For semantic reasoning and SPARQL-based querying, we utilized \texttt{GraphDB} as the backend triple store. This enabled automated consistency checking and runtime extraction of relevant ODD attributes based on environmental context.

\subsection{Confidence Model Implementation (BN)} 
\begin{figure}[!h]
    \centering
    \includegraphics[width=0.9\linewidth]{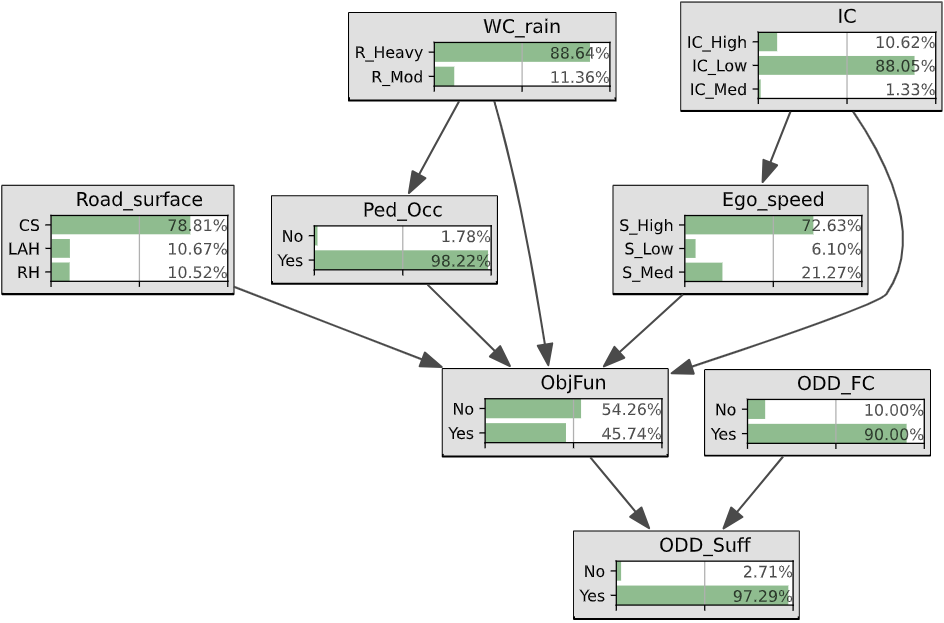}
    \caption{Example BN for \texttt{ODD\_Suff}.}
    \label{fig:BN_ODD_SUFF}
\end{figure}

To implement the BNs used for runtime confidence inference, we used \texttt{PyAgrum}, a probabilistic graphical modeling library to build and evaluate the BNs. It also allows for the computation of joint and posterior probabilities and integration of evidence (i.e., observed ODD attributes) during runtime. We illustrate in Fig. \ref{fig:BN_ODD_SUFF} an example of BN for \texttt{ODD\_Suff}. 

\subsection{Open-DSC}

\texttt{Open-DSC} is a Java-based implementation that provides an intuitive, user-friendly interface to support users across the main stages of our proposed methodology: ODD determination, safety argument formalization, and confidence model development. 

\subsubsection{ODD Determination in Open-DSC}

\begin{figure}[!h]
    \centering
    \includegraphics[width=1\linewidth]{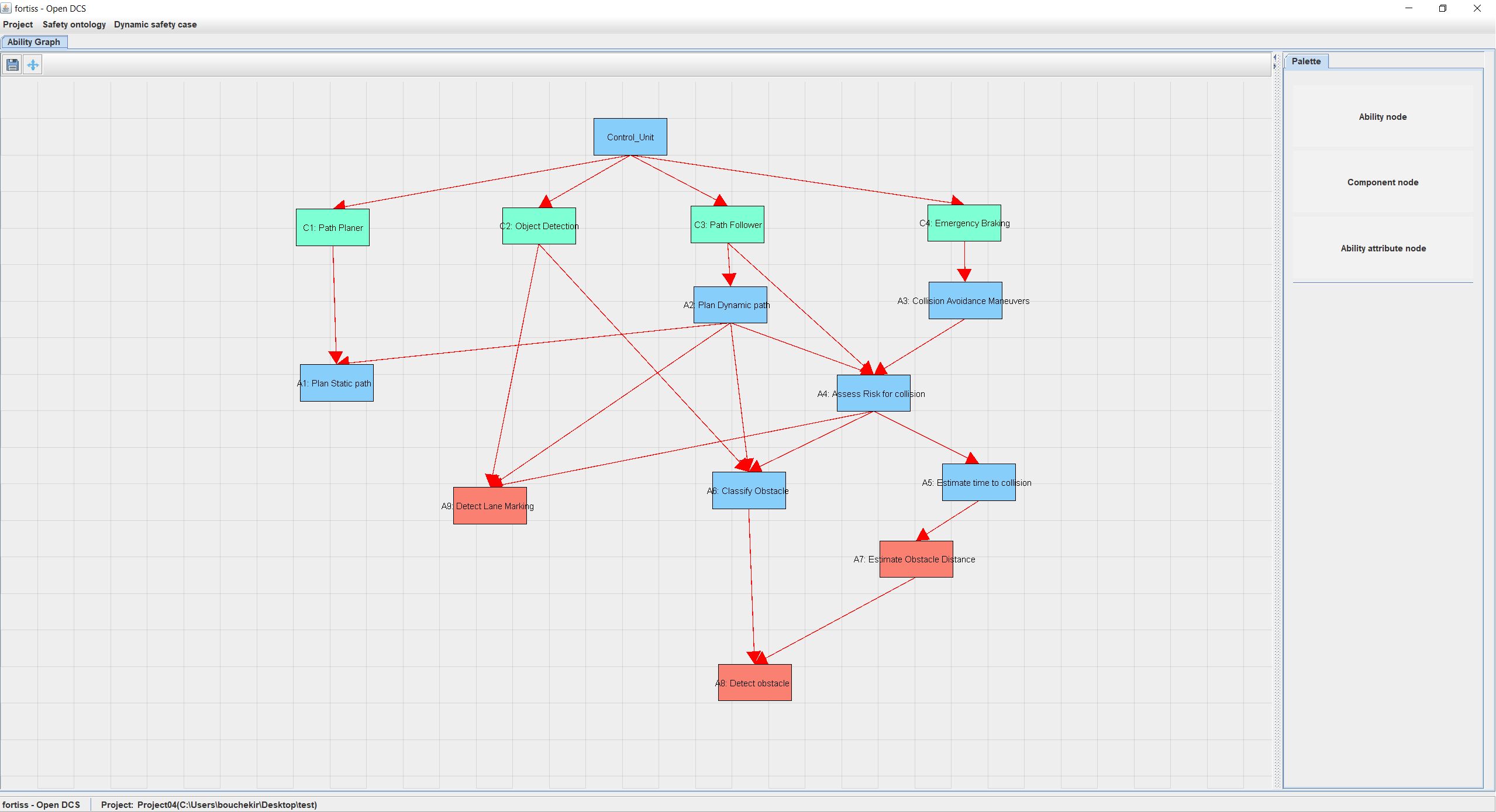}
    \caption{Determination of ODD classes in \texttt{Open-DSC} using ability graphs.}
    \label{fig:ability_graph_png}
\end{figure}

\begin{figure}[!h]
    \centering
    \includegraphics[width=1\linewidth]{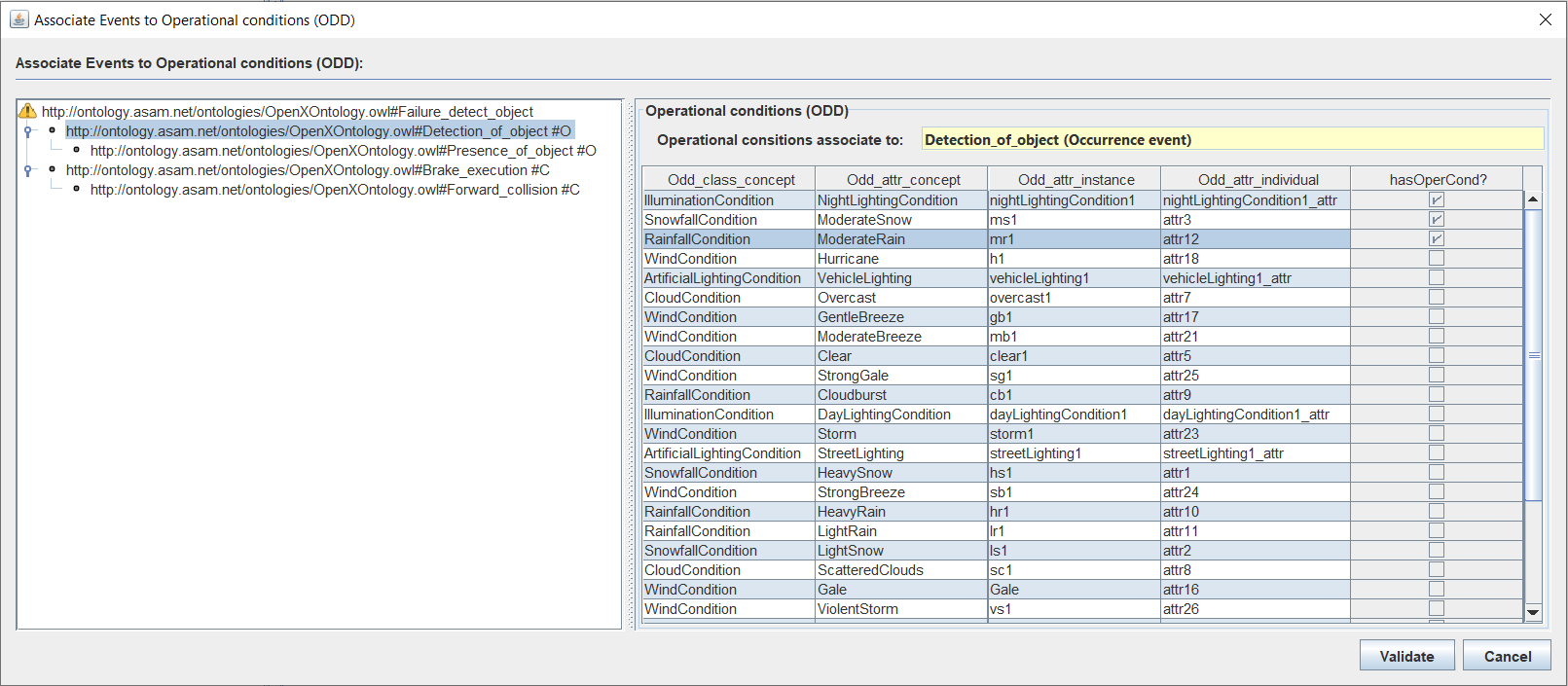}
    \caption{Risk-based derivation of ODD attributes in \texttt{Open-DSC}.}
    \label{fig:ODD_attributes_png}
\end{figure}

\texttt{Open-DSC} supports structured ODD determination through both ability graphs and risk-based causal analysis. Fig.~\ref{fig:ability_graph_png} illustrates how the tool utilizes ability graphs to identify relevant ODD classes by capturing dependencies between system capabilities and environmental factors. Fig.~\ref{fig:ODD_attributes_png} demonstrates how specific ODD attributes are derived from hazard-based causal chains, formalized through risk analysis.

\subsubsection{Safety Ontology Integration in Open-DSC}
\begin{figure}[!h]
    \centering
    \includegraphics[width=1\linewidth]{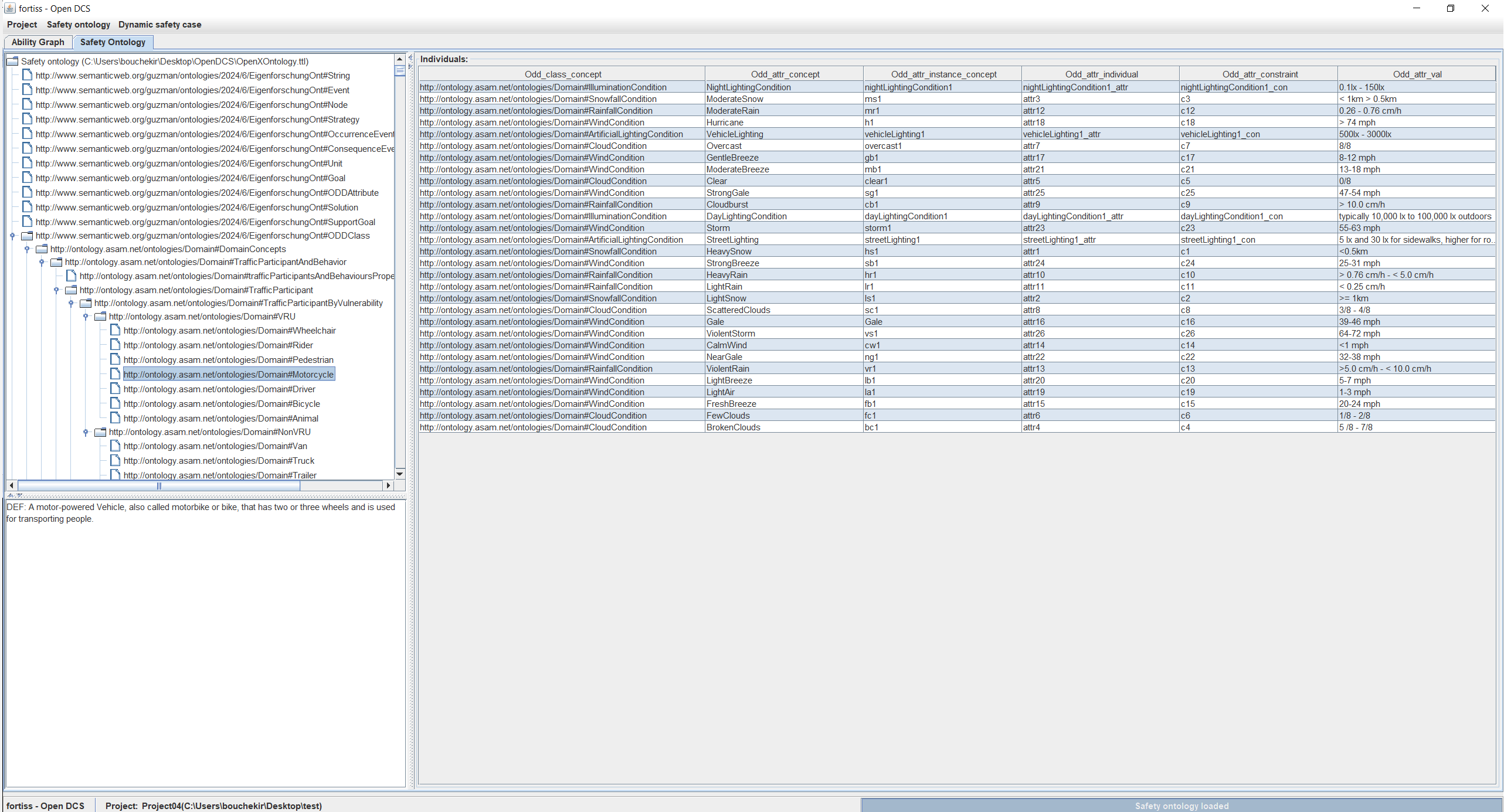}
    \caption{Visualization of the \textit{Safety Ontology} within \texttt{Open-DSC}. Users can interact with, update, and export the ontology throughout the assurance process.}
    \label{fig:OpenDCS_Safety_Ontology}
\end{figure}

\texttt{Open-DSC} adopts the \textit{Safety Ontology}\footnote{\url{https://git.fortiss.org/depai/dyn-sc/-/blob/main/OpenXOntology.ttl?ref_type=heads}} as a shared semantic foundation, enabling consistent representation of outputs at each stage. This ontology-driven framework supports advanced reasoning and SPARQL-based querying, thereby facilitating automated validation, traceability, and integration across assurance artifacts. As illustrated in Fig.~\ref{fig:OpenDCS_Safety_Ontology}, \texttt{Open-DSC} provides users with functionalities to load, modify, and export the safety ontology.

\subsubsection{Confidence Model Creation in Open-DSC}
\begin{figure}[!h]
    \centering
    \includegraphics[width=1\linewidth]{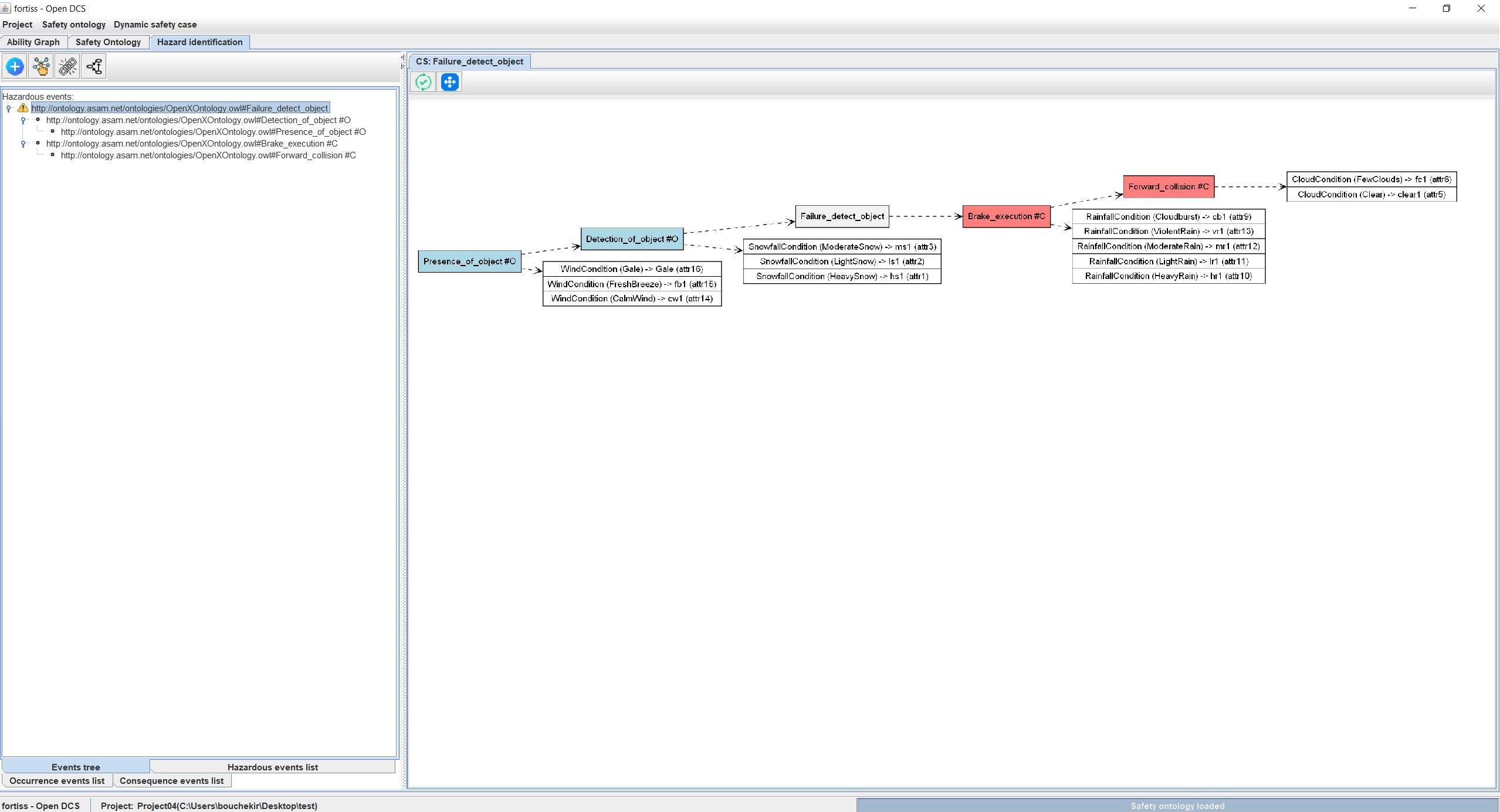}
    \caption{Causal chain of occurrence and consequence events related to a hazard, used for BN construction in \texttt{Open-DSC}.}
    \label{fig:HARA_png}
\end{figure}

\begin{figure}[!h]
    \centering
    \includegraphics[width=1\linewidth]{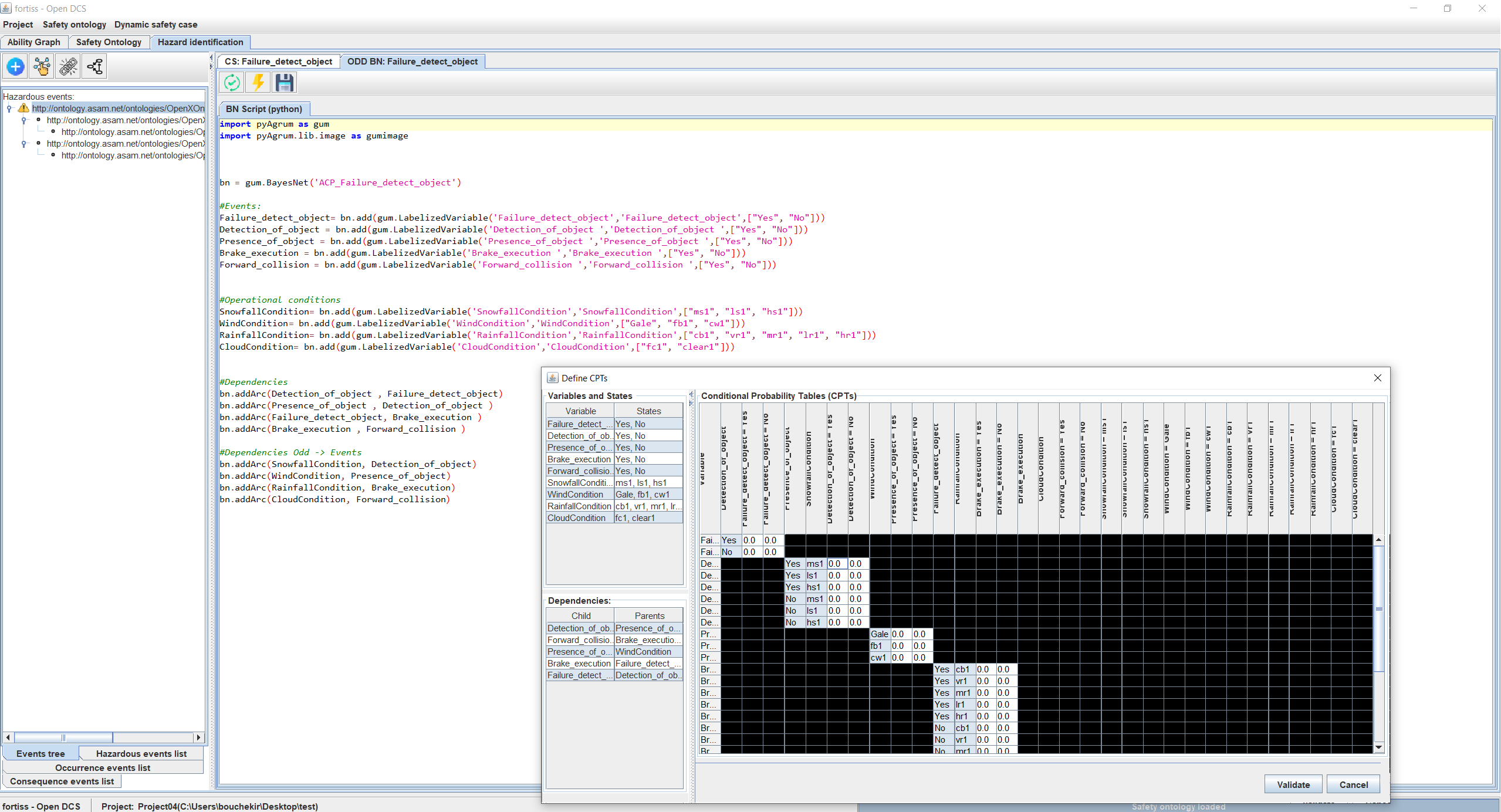}
    \caption{User interface in \texttt{Open-DSC} for updating CPTs (ACP tables) in the Bayesian Network, using expert input or simulation data.}
    \label{fig:BN_png}
\end{figure}

\texttt{Open-DSC} supports the construction of the confidence model, represented as a BN, by utilizing the causal structure derived from hazard analysis. This includes modeling the chain of occurrence and consequence events associated with specific hazards, as illustrated in Fig.~\ref{fig:HARA_png}. These causal links provide the foundational topology for the BN. Furthermore, \texttt{Open-DSC} enables users to define and update the CPTs associated with each node in the BN. This can be performed either manually, based on expert knowledge, or automatically through the analysis of simulation traces, as shown in Fig.~\ref{fig:BN_png}. 

\subsection{Simulating Runtime Behavior in CARLA} 
To demonstrate runtime safety assurance, we extended the AVP scenario in CARLA by integrating the Bayesian-based confidence model into the vehicle control loop. A perception-enabled agent (based on \texttt{automatic\_control}) was deployed with camera and LiDAR sensors. Dynamic ODD conditions were simulated through a custom script that modified weather in real time \footnote{\url{https://git.fortiss.org/depai/dyn-sc/-/tree/main/AVP_Ocluded_padestien_Simulation_CARLA?ref_type=heads}}. The BN consumed environmental parameters, such as weather, road type, and ego speed, to compute confidence scores during simulation. 

\begin{figure}[!h] 
\centering 
\includegraphics[width=\linewidth]{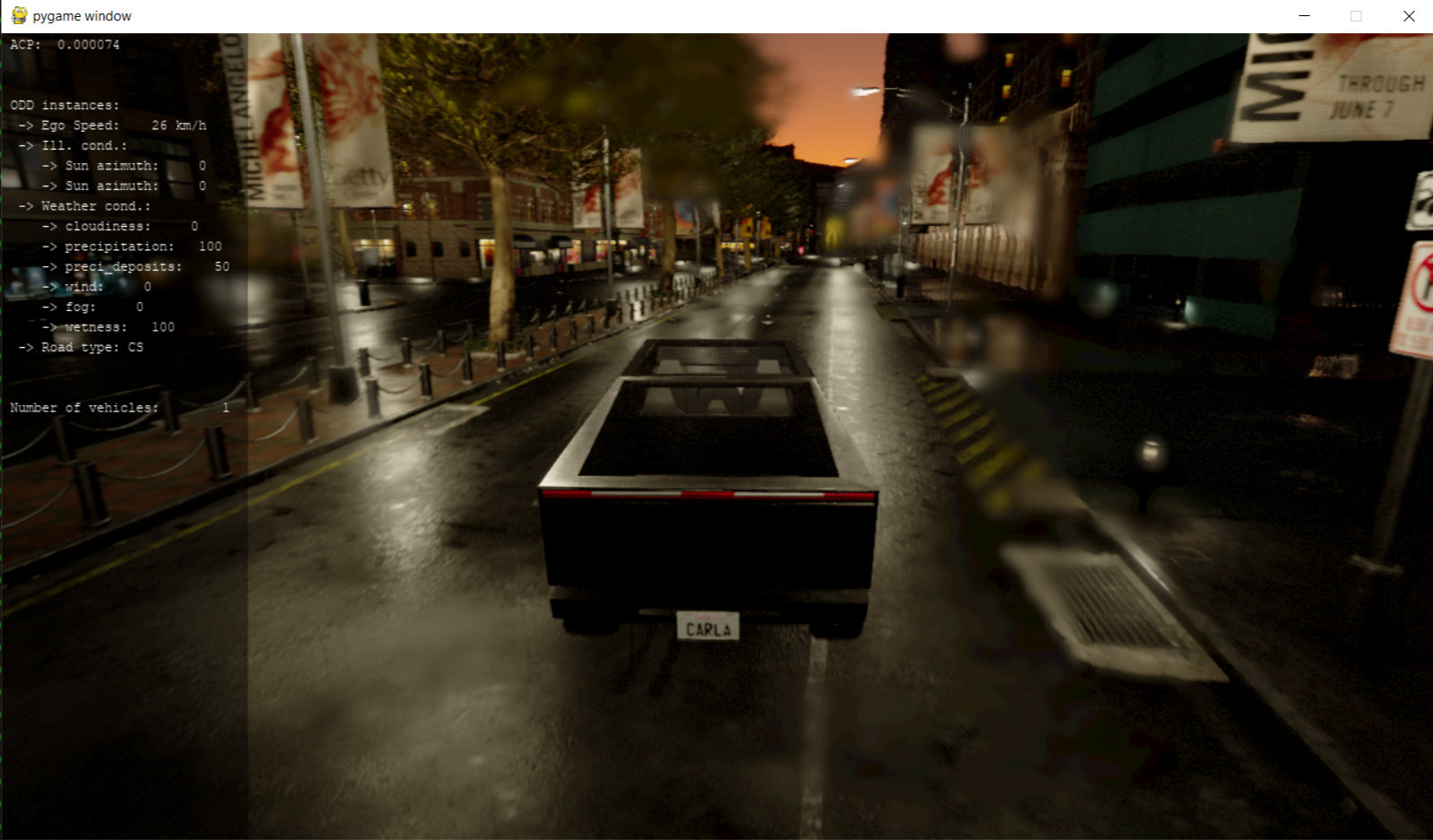} \caption{Confidence value generated using BN model.} \label{fig:ACP_Carla_01} \end{figure}

\section{Conclusion}
\label{sec:conclusion} 

In this paper, we presented a methodology for runtime safety assurance that combines formal ODD specification, structured safety argumentation, and Bayesian confidence modeling. By integrating model, data, and environmental uncertainties, our approach enables continuous, context-aware validation of autonomous system safety.

\bibliographystyle{plain}
\bibliography{ref.bib}

\begin{thebibliography}{10}

\bibitem{asam2021openodd}
{ASAM}.
\newblock {OpenODD: Concept Paper}.
\newblock \url{https://www.asam.net/standards/detail/openodd}, October 2021.
\newblock [Online].

\bibitem{BoschAVP}
Bosch.
\newblock Automated valet parking.

\bibitem{cheng2018quantitative}
Chih-Hong Cheng, Chung-Hao Huang, and Hirotoshi Yasuoka.
\newblock Quantitative projection coverage for testing ml-enabled autonomous
  systems.
\newblock In {\em Automated Technology for Verification and Analysis: 16th
  International Symposium, ATVA 2018, Los Angeles, CA, USA, October 7-10, 2018,
  Proceedings 16}, pages 126--142. Springer, 2018.

\bibitem{cho2020operational}
HongSeok Cho and R~John Hansman.
\newblock Operational design domain (odd) framework for driver-automation
  systems.
\newblock 2020.

\bibitem{automated2020avsc}
Automated Vehicle~Safety Consortium et~al.
\newblock Avsc best practice for describing an operational design domain:
  Conceptual framework and lexicon.
\newblock {\em SAE Industry Technologies Consortia}, 2020.

\bibitem{denney2015dynamic}
Ewen Denney, Ganesh Pai, and Ibrahim Habli.
\newblock Dynamic safety cases for through-life safety assurance.
\newblock In {\em 2015 IEEE/ACM 37th IEEE International Conference on Software
  Engineering}, volume~2, pages 587--590. IEEE, 2015.

\bibitem{esen2023simulationbased}
Hasan Esen and Brian Hsuan-Cheng Liao.
\newblock Simulation-based safety assurance for an avp system incorporating
  learning-enabled components, 2023.

\bibitem{gal2016dropout}
Yarin Gal and Zoubin Ghahramani.
\newblock Dropout as a bayesian approximation: Representing model uncertainty
  in deep learning.
\newblock In {\em international conference on machine learning}, pages
  1050--1059. PMLR, 2016.

\bibitem{gyllenhammar2020towards}
Magnus Gyllenhammar, Rolf Johansson, Fredrik Warg, DeJiu Chen, Hans-Martin
  Heyn, Martin Sanfridson, Jan S{\"o}derberg, Anders Thors{\'e}n, and Stig
  Ursing.
\newblock Towards an operational design domain that supports the safety
  argumentation of an automated driving system.
\newblock In {\em 10th European Congress on Embedded Real Time Systems (ERTS
  2020)}, 2020.

\bibitem{hawkins2011new}
Richard Hawkins, Tim Kelly, John Knight, and Patrick Graydon.
\newblock A new approach to creating clear safety arguments.
\newblock In {\em Advances in Systems Safety: Proceedings of the Nineteenth
  Safety-Critical Systems Symposium, Southampton, UK, 8-10th February 2011},
  pages 3--23. Springer, 2011.

\bibitem{hawkins2021guidance}
Richard Hawkins, Colin Paterson, Chiara Picardi, Yan Jia, Radu Calinescu, and
  Ibrahim Habli.
\newblock Guidance on the assurance of machine learning in autonomous systems
  (amlas).
\newblock {\em arXiv preprint arXiv:2102.01564}, 2021.

\bibitem{iso26262}
{International Organization for Standardization}.
\newblock {ISO 26262: Road vehicles – Functional safety}.
\newblock Standard, 2018.
\newblock ISO 26262:2018, Parts 1–12.

\bibitem{sotif}
{International Organization for Standardization}.
\newblock {ISO/PAS 21448: Road vehicles – Safety of the intended
  functionality (SOTIF)}.
\newblock Publicly Available Specification, 2019.
\newblock ISO/PAS 21448:2019.

\bibitem{irvine2021two}
Patrick Irvine, Xizhe Zhang, Siddartha Khastgir, Edward Schwalb, and Paul
  Jennings.
\newblock A two-level abstraction odd definition language: Part i.
\newblock In {\em 2021 IEEE International Conference on Systems, Man, and
  Cybernetics (SMC)}, pages 2614--2621. IEEE, 2021.

\bibitem{koopman2024redefining}
Philip Koopman and William Widen.
\newblock Redefining safety for autonomous vehicles.
\newblock In {\em International Conference on Computer Safety, Reliability, and
  Security}, pages 300--314. Springer, 2024.

\bibitem{lee2020identifying}
Chung~Won Lee, Nasif Nayeer, Danson~Evan Garcia, Ankur Agrawal, and Bingbing
  Liu.
\newblock Identifying the operational design domain for an automated driving
  system through assessed risk.
\newblock In {\em 2020 IEEE Intelligent Vehicles Symposium (IV)}, pages
  1317--1322. IEEE, 2020.

\bibitem{louhichi2023new}
Rim Louhichi and Insaf Sassi.
\newblock New definition and specification of operational design condition for
  autonomous railway system.
\newblock In {\em 33rd European Safety and Reliability Conference}, pages
  1918--1925. Research Publishing Services, 2023.

\bibitem{macher2015sahara}
Georg Macher, Harald Sporer, Reinhard Berlach, Eric Armengaud, and Christian
  Kreiner.
\newblock Sahara: a security-aware hazard and risk analysis method.
\newblock In {\em 2015 Design, Automation \& Test in Europe Conference \&
  Exhibition (DATE)}, pages 621--624. IEEE, 2015.

\bibitem{reschka2015ability}
Andreas Reschka, Gerrit Bagschik, Simon Ulbrich, Marcus Nolte, and Markus
  Maurer.
\newblock Ability and skill graphs for system modeling, online monitoring, and
  decision support for vehicle guidance systems.
\newblock In {\em 2015 Ieee intelligent vehicles symposium (Iv)}, pages
  933--939. IEEE, 2015.

\bibitem{shakeri2024operational}
Ali Shakeri.
\newblock Operational design domains in automated vehicles: A review of
  state-of-the-art standards, challenges, and proposed solution.
\newblock 2024.

\bibitem{shih2018formal}
Andy Shih, Arthur Choi, and Adnan Darwiche.
\newblock Formal verification of bayesian network classifiers.
\newblock In {\em International Conference on Probabilistic Graphical Models},
  pages 427--438. PMLR, 2018.

\bibitem{sorokin2024towards}
Lev Sorokin, Radouane Bouchekir, Tewodros~A Beyene, Brian Hsuan-Cheng Liao, and
  Adam Molin.
\newblock Towards continuous assurance case creation for ads with the
  evidential tool bus.
\newblock In {\em European Dependable Computing Conference}, pages 49--61.
  Springer, 2024.

\bibitem{sun2021acclimatizing}
Chen Sun, Zejian Deng, Wenbo Chu, Shen Li, and Dongpu Cao.
\newblock Acclimatizing the operational design domain for autonomous driving
  systems.
\newblock {\em IEEE Intelligent Transportation Systems Magazine}, 14(2):10--24,
  2021.

\bibitem{wozniak2020safety}
Ernest Wozniak, Carmen C{\^a}rlan, Esra Acar-Celik, and Henrik~J Putzer.
\newblock A safety case pattern for systems with machine learning components.
\newblock In {\em Computer Safety, Reliability, and Security. SAFECOMP 2020
  Workshops: DECSoS 2020, DepDevOps 2020, USDAI 2020, and WAISE 2020, Lisbon,
  Portugal, September 15, 2020, Proceedings 39}, pages 370--382. Springer,
  2020.

\end{thebibliography}

\end{document}